\documentstyle[11pt,aasms4]{article}

\def\plotfiddle#1#2#3#4#5#6#7{\centering \leavevmode
    \vbox to#2{\rule{0pt}{#2}}
    \includegraphics{#1}}

\def\sun{\hbox{$\odot$}}

\lefthead{}
\righthead{}

\begin{document}
\title{Palomar 13's Last Stand}

\author{M. H. Siegel\altaffilmark{1}, S. R. Majewski\altaffilmark{1,2,3},
\\K. M. Cudworth\altaffilmark{4} and  M. Takamiya\altaffilmark{5}}

\altaffiltext{1}{Dept. of Astronomy, University of Virginia, P.O. Box 3818, University Station,
Charlottesville, VA, 22903-0818 (mhs4p@virginia.edu,srm4n@didjeridu.astro.virginia.edu)}

\altaffiltext{2}{David and Lucile Packard Foundation Fellow, Cottrell Scholar}

\altaffiltext{3}{Visiting Research Associate, The Observatories of the Carnegie
Institution of Washington, 813 Santa Barbara Street, Pasadena, CA 91101}

\altaffiltext{4}{Yerkes Observatory, University of Chicago, 73 W. Geneva St., Williams Bay, WI, 53191 
(kmc@hale.yerkes.uchicago.edu)}

\altaffiltext{5}{Gemini Observatory, 670 N. A'ohoku Place, Hilo, HI, 96720 (mtakamiya@gemini.edu)}

\begin{abstract}

	We present a proper motion and CCD photometric study of stars in the distant halo globular cluster Palomar 13.
The absolute proper motion of Pal 13 with respect to the background galaxies, derived
from moderate scale photographic plates separated by a 40-year baseline, is
$(\mu_{\alpha cos \delta}, \mu_{\delta}) = (+2.30, +0.27) \pm (0.26, 0.25)$ milliarc-seconds per year.  The resultant total space 
velocity (315 km s$^{-1}$) implies that Pal 13 is in the inner part of its orbit near perigalacticon.
Orbital integration reveals the cluster to possess 
an inclined, very eccentric, retrograde orbit.  These data confirm that Pal 13 is a paradigm ``young halo"
globular cluster.

The derived proper motions for cluster stars are used to produce membership probabilities and a cleaned CCD UBV
catalogue for Pal 13.  With this data set we have made small revisions to Pal 13's distance, metallicity,
position and light profile.  The membership of four previously reported RR Lyrae variables and a 
proportionally large group of blue straggler stars are confirmed.  As expected, the blue stragglers are centrally 
concentrated.  

The small size of this cluster, combined with the shape of its light profile, which shows a clear departure
from a classical King function beyond the tidal radius, suggests that Pal 13 is in the final throes of destruction. 
This could explain the large blue straggler specific frequency, as destructive processes 
would preferentially strip less massive stars.

\end{abstract}


\section{Introduction}

	The globular cluster Palomar 13 is one of a number of faint, low mass, sparse globular clusters identified from the 
Palomar Observatory Sky Survey (Abel 1955; Wilson 1955).  The first study of this cluster by Ciatti et al. (1965) identified 
four RR Lyrae stars in the cluster and derived a distance modulus of $(m-M)_{V} =17.11$.  At  
$(\alpha, \delta)_{1950}=(23:04.2, +12:28)$, 
or $(l,b)=(87.1^{\circ}, -42.7^{\circ}$), this distance
modulus places Pal 13 at a Galactocentric radius of 27.2 kpc, well into the halo of the Milky Way.

Further information on Pal 13 did not appear in the literature for over a decade when a flurry of spectroscopic
studies of Pal 13 revealed it be moderately metal poor.  Spectroscopic [Fe/H] were measured
at $-1.9 \pm 0.4$ (Canterna \& Schommer 1978), $-1.75 \pm 0.6$ (McClure \& Hesser 1981), $-1.67 \pm 0.15$
(Zinn \& Diaz 1982) and $-1.9 \pm 0.1$ (Friel et al. 1982).  In the most comprehensive study 
to date, Ortolani et al. (1985, hereafter ORS) reexamined the 200-inch Palomar plates that were the basis for the 
Ciatti et al. study (as well as the present astrometric work).  ORS confirmed the distance modulus, 
determined a metallicity between [Fe/H]=-1 and -1.5, and found 
that Pal 13 was the faintest and smallest globular cluster measured at that time\footnote{Harris (1996) 
currently lists Pal 13 as the fifth smallest globular cluster based upon its luminosity.}.  They claimed that Pal 13's
luminosity function is similar to that 
of the bright globular cluster M3, but scaled down by a factor of $\sim 60$.  From this, ORS estimated an absolute magnitude of 
$M_{V}$=-3.36 and a mass of approximately 3 x $10^3$ solar masses with an assumed (M/L) ratio of 1.6.

	Recently, the CCD photometry of Borissova et al. (1997, hereafter B97) revealed Pal 13 to have an age of 
approximately 12 Gyr based on the fitting of isochrones from Proffitt \& VandenBerg (1991) and Bergbusch \& 
VandenBerg (1992) - suggesting that it is representative of the ``young halo" population of globular clusters.  In addition, B97 
reported a number of possible blue straggler stars, a Zinn-West (1984) metallicity of [Fe/H]=-1.52 and confirmed the 
Ciatti et al. distance modulus.

	We have undertaken a comprehensive photometric and astrometric study of Pal 13 to improve understanding 
of this object.  The present work updates and supercedes our previous report (Cudworth et al. 1993).  Our efforts have been 
directed toward three goals:

(1) The measurement of Pal 13's absolute proper motion and subsequent determination of its true space velocity.  This 
will improve our knowledge of Pal 13's orbit, the Galactic mass enclosed by that orbit and provide insights 
into its dynamical history.

(2) The production of an improved UBV CCD photometric catalogue of the Pal 13 field.

(3) The use of proper motions to separate true members of Pal 13 from the field star population.  A
cleaned color-magnitude diagram (CMD) will help to verify the membership of blue stragglers 
and RR Lyrae variables.  It will also improve our ability to compare isochrones to confirm Pal 13's
age and metallicity.
This is particularly important for Pal 13, which is a sparse cluster and does not show 
a strong contrast against the field stars except in its core.

	This is the fifth object and third globular cluster in our program (Majewski \& Cudworth 1993) 
of measuring absolute proper motions of distant globular clusters and dwarf spheroidal galaxies 
using faint galaxies and QSO's to fix the proper motion zero point (Schweitzer et al. 1995; Dinescu et al. 2000; 
Schweitzer et al. 2000; Cudworth et al. 2000).  \S2 describes our observations and \S3 our reduction 
methods.  Our photometric and astrometric results are presented in \S4 and \S5 respectively and the implications of our
work are discussed in \S6.

\section{Observations}

\subsection{Photographic Data}

Our photographic collection included 19 plates ranging in epoch from 1951 to 1991.  Table 1 lists the plate material, 
along with epochs, scales, emulsions, filters and passbands.  The plate number prefixes indicate the telescope:  
PH, Palomar Hale 200 inch and CD, Las Campanas Du Pont 100 inch.  Most plates could be measured reliably to a depth of $V\sim21$ or 
$B\sim21.5$, which barely reaches Pal 13's main sequence turn-off (MSTO) which ORS found to lie near 
$V = 21.1$.

Stars and compact galaxies were selected for measurement within a radius $\sim8\farcm0$ from our initial estimate of the 
cluster center.  This limit was set by the coma-free field of the Hale plates.  A total of 595 centroidable objects were 
within this radius.  This area is much larger than the cluster 
itself, measured to have a tidal radius of 3\farcm2 by Harris \&
Racine (1979), 3\farcm8 by ORS and 2\farcm2 by Trager et al. (1995).  
Our choice of a larger area to scan was motivated by our need to get as many members as possible in this sparse cluster, 
to look for stars evaporated and/or stripped from the cluster and to include as many galaxies as possible to 
set the proper motion zero point, the latter being the fundamental limitation on the accuracy of absolute
proper motion measures.  The area initially selected would later prove to be substantially 
larger than the astrometrically usable field.

The plates were scanned with the PDS microdensitometer at MADRAF (Midwest Astronomical Data Reduction and Analysis Facility) in 
Madison, WI.  Scanning and centroiding procedures closely followed those described by Cudworth (1985, 1986).

\subsection{CCD Data}

We observed Pal 13 on UT 19-23 July and 31 July-2 August, 1991 with the Las Campanas 1-m Swope telescope using the 
thinned $1024^2$ Tek 2 CCD camera.  After dithering between $UBV$ sets of 
observations in order to increase sky coverage and remove cosmetic defects in the chip, a total of seven 
images in Johnson $V$ (300-900 seconds),
five images in Johnson $B$ ( 900-1800 seconds) and four images in Johnson $U$ (1800-2400 seconds) were
obtained.  Six of 
the images were taken under non-photometric conditions.  Seeing ranged from 1\farcs6 to 2\farcs4 full width half maximum with a 
mean of 2\farcs0.  The total spatial coverage of the dithered frames was of an area $\sim$16' on a side.  When the images were 
combined, they provided deeper coverage of an area 10\farcm6 on each side, comparable to the 10\farcm9 wide area of the 
astrometric field.  Only 14 objects for which proper 
motions were measured failed to appear in the eventual CCD catalogue.

\section{Reductions}

\subsection{CCD Photometry}

The CCD data were reduced to the flat field stage with the IRAF\footnote{IRAF is distributed by the National 
Optical Astronomy Observatories,
which are operated by the Association of Universities for Research in 
Astronomy, Inc., under cooperative agreement with the National Science Foundation.} package CCDRED.
All CCD frames were photometered with the DAOPHOT point-spread function (PSF) photometry 
program (Stetson 1987) using a geometrically variable 
Moffat point-spread function.  This photometry was then input to the ALLFRAME program (Stetson 1994), which solves 
magnitudes and positions on all frames in the data set to produce a consistent result for every image.  This improved 
the photometric precision while extending it to fainter magnitudes in the area overlapped by multiple CCD images.

The detections were matched using DAOMASTER and then calibrated to observed Graham (1982) standard 
stars using procedures described in Siegel \& Majewski (2000) that account for frame-to-frame residuals.
Typical frame-to-frame 
residuals were on the order of 0.01 magnitudes, with a maximum of 0.03 magnitudes among the photometric 
observations.  Non-photometric observations naturally had very large residuals (up to 0.9 magnitudes in the initial 
average) which were evaluated and removed by comparison to the photometric frames.
The resulting photometric precision is $(\sigma_V,\sigma_B,\sigma_U) = (0.02,0.03,0.06)$ at $V=20$.

We have compared our photometric results to the previous efforts of ORS and B97 (Figure 1) and have discovered some
discrepancies.  The ORS survey, which used photographic photometry calibrated to a photoelectic sequence
in the field, is very close to our own result in the $V$ passband.  However, there is a non-linearity
in the $B$ band comparison.  We believe the non-linearity in our comparison to ORS to be the result of innaccurate 
faint $B$ magnitudes in the photoelectric sequence used for that study.  In particular, while the $V$ magnitudes of 
the faint stars in the ORS photoelectric sequence are similar to our own, the colors of 
the two faintest are off by 0.1 magnitudes.  This produces a non-linear effect effect in a sequence that was defined by only 
five $B$ magnitude photoelectric measurements.

The comparison of our photometric data to that of B97 shows a large scatter but no systematic effects.  B97 also
reported significant scatter in their comparison to ORS of ($\sigma_V,\sigma_{B-V}) = (0.18, 0.09)$.  On the other
hand, our comparison to ORS shows small scatter, but systematic deviations at fainter magnitudes.  We interpret
figure 1 to suggest that our photometry matches the superior precision of ORS while being free of their systematics.

\subsection{Photographic Photometry}

We utilize photographic colors and magnitudes for all stars for which proper motions are to be obtained in order to 
derive and remove color- and magnitude-dependent systematic effects in the astrometry, effects that can be exacerbated
by combining data from different telescopes, filter systems, cameras, types of plates and, in this case, different
hemispheres.
\footnote{A similar reduction with the same kind of plate material for the globular cluster Pal 12 
(Dinescu et al. 2000) found very strong color terms in the plate solutions.  This was the result of combining data 
from different hemispheres.  In particular, 
astrometry with the Hale 200" telescope was very sensitive to color terms induced by observing
at zenith angles that were necessarily $\geq 54^{\circ}$.  
The effect is present in our study, but lessened as the zenith angle of Pal 13 is
$\geq 41^{\circ}$ from Las Campanas.}
Thus the PDS data were 
used to derive photographic $V$ and $B-V$ for each star in our sample 
using the method and software described by Cudworth (1985, 1986).  Substituting photoelectric photometry would be inappropriate
as these systematic effects are dependent upon the color and magnitude at the time of observation.

As is generally the case with photographic photometry, a calibration using photoelectric or CCD data was required.
Our first photometric reduction used the photoelectric sequence of ORS.  We found this sequence to produce high 
non-linearity in the $B$ photometry of the faintest objects, manifested by an unrealistic blueward curve at the 
faint end of the giant branch and MSTO.  After this initial reduction, the new CCD 
photometry we obtained (discussed above) was used to correct our photographic photometry with a substantial 
subsequent improvement in the precision and accuracy of the astrometry.
Random photometric errors in our photographic ($V, B-V$) ranged from $\pm (0.015, 0.020)$ for $V \lesssim 18.5$ to 
$\pm (0.10, 0.12)$ near the magnitude limit of $V \sim 21$.

\subsection{Proper Motions}

We derived proper motions from PDS stellar centroid measures using the latest revision of the central-overlap software 
described in its initial version by Cudworth (1985, 1986) and subsequently updated in Peterson \& Cudworth (1994)
and Cudworth et al. (2000).  A brief description follows.

All 19 plates were included in the solution.  The plate 
constants included linear and quadratic terms in coordinates ($x,y$) and $V$ magnitude ($m$), coma ($mx$,$my$), 
$B-V$ color ($c$) and color magnification ($cx$,$cy$).

All plates required corrections for distortion.  An iterative solution similar to that described by 
Murray (1971) was used to find an accurate distortion center for each plate.  Nominal values of the distortion 
constants for the Du Pont plates (Cudworth \& Rees 1991) were used to remove most of the distortion 
from our selected ``standard" plate, CD-2947.  This was our best plate from the Las Campanas 100-inch telescope (which has a 
small distortion, especially over the small field we were working in).  Distortion terms ($xr^2$ on Du Pont plates 
and both $xr^2$ and $xr^4$ terms on the Hale plates, where $r = (x^2+y^2)^{\frac{1}{2}}$ is the radial 
distance from the distortion center, and where similar $y$ terms were used in the $y$ solution) were then 
included among the plate constants of all remaining plates to account for any differences between the distortion 
of an individual plate and the undistorted coordinate system of CD-2947.  The coma in the 200-inch Palomar 
plates is very strong.  To reduce this effect, we were eventually forced to restrict our analysis to objects 
within 30 mm (5\farcm5) of the cluster center.  Our most shallow plate (CD-2954) did 
not have enough stars to constrain all of the plate constants.  We therefore removed two terms for which the error 
was larger than the coefficient itself.

Galaxy images were treated exactly like stars in these reductions but were 
excluded from the list of stars from which the plate constant solutions were generated.
Preliminary reductions were also used to identify and remove poor measurements due to crowding, poor image quality or 
plate flaws.  This selection, in combination with the reduced field size necessitated by the Palomar coma, 
left us with 265 objects in the Pal 13 field for which proper motions could be measured.

Although the software we used is fundamentally the same as that described initially
by Cudworth (1985, 1986), the Pal 13 reductions differed in two significant ways:  (1) we 
iterated to find the most accurate assessment of the location of the optical axis before solving for the 
plate constants, as noted above; and (2) we 
removed stars with large proper motions from the list of stars 
used for the plate constant solutions.  In the final reductions, 
photometry and preliminary proper motions were used to define a list of stars composed as purely as possible
of cluster members.  These cluster members were the only stars used to determine the plate constants.
This has the effect of separating observational systematic effects 
(distortion, refraction, etc.) from secular terms (differences in proper motion), producing more accurate plate 
constants because these constants are defined by stars of identical proper motion.  The cluster members cover a 
large enough portion of the astrometric field to provide good constraints upon geometric constants.  They also 
span a large range in magnitude ($17 < V < 22$) and thus provide good constraints upon magnitude terms.  However, 
one remaining concern lies with the color terms.  The cluster stars only cover a small range in 
color ($0.16 < B-V < 0.88$).  While color terms may be well-constrained within this range, the lack of constraints 
outside of it could cause extremely blue or extremely red objects to have systematic proper motion errors, 
including any red or blue extra-galactic objects used to set the absolute zero point of the proper motion (see \S5.2).

The resulting proper motions are in a relative system with the zero point 
defined by the mean motion of the pre-defined cluster members.  We have carefully examined plots of 
the final proper motions against coordinates, magnitude and color to search for any possible 
systematic trends (see Figure 2).  No such trends were found.  In particular, we do not find evidence in these 
motions for systematic problems due to the PDS machine such as we found and removed in our study of the Ursa Minor 
dSph (Schweitzer et al. 2000).  We also looked for PDS-related systematic trends in the residuals from the plate 
constant solutions for individual plates, but even on the best plates random scatter masked any clearly 
significant systematic effects beyond those removed by the plate constants.

\subsection{Cluster Membership}

Cluster membership probabilities were derived from the proper motions using a recent revision of the standard 
Yerkes probability software based on the ideas discussed in detail by Dinescu et al. (1996) in their study of 
NGC 188 (though a similar technique was also used by Stetson 1980).  The key difference between this study and previous 
Yerkes cluster studies is that the measurement errors used to define the probabilities ($\sigma_{xc}$ and $\sigma_{yc}$)
are the errors of the {\it individual stellar proper motions}.  Previously, probabilities were based upon the difference between
the proper motion of an individual star and that of the cluster mean, scaled by
the Gaussian disperison of the cluster proper motion distribution.  The parameters describing the 
cluster and field star Gaussian distributions are listed in Table 2.  Here $N$ is the number of stars in each 
distribution; $\mu_{x0}$ and $\mu_{y0}$ are the centers of
the distributions, and $\sigma_{x0}$ and $\sigma_{y0}$ are the dispersions of the distributions in milliarc-seconds (mas) per year.  We repeat
that the numbers for $\sigma_{x0}$ and $\sigma_{y0}$
were not used in the final probability derivation for each star and are tabulated here only to 
allow for comparison with work in other clusters.  

Note that these membership probabilities are based solely upon proper motion.  Obviously a star
that is located near the red giant branch in color-magnitude space is more likely to be a member than one of equal
astrometric probability that is far removed from the obvious and likely sequence of cluster stars.  However, we have elected
to take no such photometric consideration into account in our probability derivation because it might exclude
unusual stars or stars in a short-lived stage of stellar evolution.  The astrometric
selection requires fewer {\it a priori} assumptions.

\subsection{Positions}

The relative positions in our initial catalogue were 
transformed to the system of the USNO-A2 catalogue (Monet et al. 1996), using 96 stars
in common between USNO-A2 and our catalogue.  This
transformation shows a global disperson of 0\farcs3.  Relative positions are from our astrometric catalogue and
are accurate to within $\sim$0\farcs01.

\subsection{The Catalogue}

The final catalogue for Pal 13 consists of 421 objects that either have a measured proper motion or CCD
$B$ and $V$ photometry errors less than 0.15 magnitudes.
Table 3 lists the identification number in our study, the cross-identification
number in ORS and B97 and the
CCD magnitudes and errors in $UBV$.  The objects are sorted by $V$ magnitude.  Table 4 lists the 
equatorial coordinates, relative proper motions and the proper motion errors in units of mas yr$^{-1}$.  The final column 
of Table 4 lists the astrometric membership 
probabilities $P$ in percent.  $P$ = -1 indicates galaxies, which are listed at the top of the tables; their 
photometry may be less reliable than that for stars. Values listed as ``..." were not measurable.  
The first pages of the tables are printed here, with the full tables available electronically.

\section{Photometric Properties}

\subsection{The Red Giant Branch and Metallicity}

The $(B-V,V)$ CMD of Pal 13 is shown in Figure 3a.  The photometry is 
from our CCD data set and has been corrected for reddening using a value of $E_{B-V}=0.11$ from
Schlegel et al. (1998).  Figure 3b shows the same CMD with proper motion data included.
The most apparent characteristic of Pal 13's CMD is its extreme sparseness.  Only seven probable ($P \ge 50\%$) members 
are brighter than $V=18.5$.  Even allowing stars with a probability level as low as 20\% only 
adds another three stars to this magnitude range.  Only two stars with $P \ge 20\%$ are brighter than the 
horizontal branch (ID's 22 and 25), although 
a few bright giants may have fallen out of the catalogue due to the limited field or crowding.  Indeed, two stars 
in our photometric catalogue (ID's 4 and 16) were too far away from the cluster center for astrometry but could belong to 
the red giant branch.

Photometric attempts to measure the metallicity of Pal 13 have already been made by ORS and B97.  Metallicity 
estimates are often made from the color of the giant branch at the magnitude of the horizontal branch and
the extreme poorness of the giant branch limits the effectiveness of such measures for this particular cluster.
We have found only two giant stars 
near this magnitude level with P$\ge 50\%$.  Their average ($B-V$) color (0.89 $\pm$ 0.03), corrected
by the cluster reddening produces a metallicity of [Fe/H]=-1.65 $\pm 0.13$ on the Zinn-West (1984) scale.
This value is close to the spectroscopic estimates mentioned above and similar to the metallicity
(-1.6 dex) derived from the period shift of the RR Lyrae stars given in ORS.  We also note that several of 
the isochrone fits in B97 converged at a metallicity of approximately [Fe/H]=-1.65.  Because of the sparseness of 
Pal 13's giant branch and the extreme sensitivity of {\it any} photometric measure of metallicity to zero point and reddening, 
a more accurate metallicity evaluation must await high resolution spectroscopy of the few Pal 13 giants.

The subgiant branch of Pal 13 is curious in that it shows an apparent parallel sequence - possibly from near equal
mass binary stars.  
Such a parallel double giant branch is also seen in the old open cluster Melotte 66, for example (c.f. Majewski et al. 2000a).  
The prominent double sequence may have a similar origin as the blue stragglers (\S 4.3) in the cluster:  stars or star systems
with more mass, possibly from merged or unmerged binaries, are more likely to be retained in a cluster undergoing evaporation 
or severe tidal disruption (see \S 6.4).  However, unlike the blue stragglers discussed below, the stars in the second subgiant branch are {\it not}
centrally concentrated.

We made several attempts at using isochrones to measure the age of Pal 13.  Unfortunately, our photometry
rapidly degrades near the MSTO of the cluster, prohibiting an accurate fit to the data.
Figure 3c shows the proper-motion cleaned CMD overlayed by an isochrone from VandenBerg \& Bergbusch (2000).  This
isochrone has [Fe/H]=-1.61, [$\alpha$/Fe]=+0.3, Y=0.236, $E_{B-V}$=0.08, an age of 12 Gyr and $m-M$=16.97 (as 
revised in \S4.4).  The latter reddening was found to produce a more consistent fit than the Schlegel et al.
value.  We compare to the lower of the two subgiant branches, which we assume to represent the location of 
stars that are not near-equal mass binaries.

While the isochrone appears to overlap the data reasonably well, it must be stressed
that this isochrone does {\it not} represent a fit to the data.  It has been constructed from the age derived
in B97 and our slight revisions to the metallicity, reddening and distance.  It is included simply for comparison.  Deeper 
photometry is needed to provide stronger constraints upon Pal 13's properties.

\subsection{Variable Stars}

We have used a modified version of the Welch-Stetson (1993) technique to detect variable stars in our CCD 
data.  We have identified four variable stars, all of which correspond to the RR Lyrae stars identified in
ORS.  Cross-identifications, CCD 
photometry and membership probabilities are listed in Table 5 with the prefix ORS-V.  
Three of these variables are likely members, including V4, which is 5' from 
the cluster center.  The proper motion of the fourth ORS variable, V2, could not be measured because of 
image crowding.

Photometry that we present for the variable stars is simply averaged over our CCD images.  We do not have adequate phase 
coverage to derive periods, amplitudes or intensity-weighted mean magnitudes for the variables.  Pal 13 has
the highest specific frequency of RR Lyrae variables (158) as a function of magnitude for any cluster in the Milky Way
(Harris 1996).

\subsection{Blue Stragglers}

We have identified seven objects 1.5 magnitudes brighter than the MSTO that are probable members of Pal 
13 ($P \ge 50\%$).  Six of these stars are within 40" of the cluster center, a radius that encloses
only 40\% of the probable cluster members.  This agrees with the finding of B97 that the blue stragglers stars (BSS) are 
centrally concentrated.

We have cross-identified the blue straggler stars listed in B97 in Table 5 with the prefix B97-BSS.  All of
the B97 blue stragglers except numbers 3 and 4 appear as blue stragglers in our study.  While star 3 is a very likely proper 
motion member and star 4 could be a member, both our CCD and photographic photometry place each of these stars 
within the MSTO.  Our additional two blue stragglers are listed with the prefix BSS.  Our result confirms B97's classification 
of Pal 13 as BS1 (second-parameter, high specific frequency of BSS, low concentration parameter) on the 
system of Fusi-Pecci et al. (1992).

\subsection{Cluster Distance}

It is important to have the most accurate distance modulus for Pal 13 that can be derived.  Analysis of its kinematics
(\S 6.2) can be dramatically altered by even small changes in the cluster's distance.
The only standard candles available in Pal 13 are 
the RR Lyrae variables.  While we lack the phase coverage in either photographic or CCD data to produce intensity-weighted 
mean magnitudes, the average magnitudes are tightly clumped - their scatter is smaller than the 
uncertainty in RR Lyrae absolute magnitudes (see Smith 1995 for discussion).  This uncertainty is amplifed when one convolves 
it with the wide spread in measured Pal 13 metallicities which translates to a spread in 
estimated $M_V(RR)$.

We find that our four RR Lyrae variables and single HB star have an average magnitude of $V=17.75$ in both photographic and CCD 
photometry.  Using the RR Lyrae absolute magnitude relation of Sandage (1993), assuming a metallicity of [Fe/H]=-1.65 and using 
the Schlegel et al. reddening, we derive a true distance modulus to Pal 13 of $(m-M)_V$=16.97.  The revised distance of Pal 13
is 24.8 kpc from the Sun, 25.8 kpc from the Galactic Center (assuming $R_{\sun}$=8.0 kpc).  The uncertainty in
this distance is dominated by the scatter in RR Lyrae absolute magnitudes ($\sim 0.2$ magnitudes) which corresponds 
to a distance uncertainty of 2.1 kpc.

\subsection{The Two-Color Diagram}

The two-color diagram of Pal 13 proper-motion selected stars is shown in Figure 4a.  While the U-band photometry
is noticeably poor, the diminutive giant sequence can be seen among the high-probability stars, stretching from $(B-V,U-B) = 
(0.4,-0.3)$ to $(B-V,U-B) = (0.8,0.2)$.  The doubling of the subgiant branch noted in \S4.1 is not apparent in the 
two-color diagram, although it could be masked by high scatter and small numbers.

$UBV$ photometry can be useful for the derivation of metallicity and reddening.  However, in the case of Pal 13 we are 
limited to using giant stars for this derivation as the dwarf stars are too faint.
The extreme paucity of 
Pal 13 giants, especially Pal 13 giants with good U-band photometry ($\sigma_U \leq 0.1$) makes this is a difficult
proposition at best.  In addition, while the ultraviolet excess is reasonably well-calibrated for main sequence dwarf
stars (c.f. Laird, Carney \& Latham 1988), no equivalent relation has been produced from CCD
photometry of giant stars.  Our attempts to calibrate giant star $UV$ excess from the literature were thwarted because
$U$-band photometry was found to be inconsistent between different
studies, possibly because of different standard stars.  Finally, metallicity estimates from $UV$ excess
are extremely sensitive to errors in the zero point and/or reddening.  A zero point error as small as 0.01 magnitudes can translate to an 
inferred metallicity error of 0.1 dex at the metal-poor end (c.f. Laird et al. 1998).

To make a simple comparison, we have overlaid a fit to the synthetic photometry 
derived by Majewski (1992) for Gunn-Stryker (1983) spectrophotometric giant stars (Figure 4b), restricting our analysis to
Pal 13 stars ($P \ge 40\%$) with good photometry ($\sigma_U \leq 0.1$).  Only the brighter Pal 13 giants have small enough errors to be 
included.  Note the clear 
ultraviolet excess of the Pal 13 giant stars and the non-linearity of this excess with effective temperature.  
We have also made a comparison to our high quality UBV photometry of giants in the metal-poor globular cluster NGC2419 
(presented in Siegel et al. 1999).  The high quality Pal 13 data appear to be located between the solar metallicity
locus of Gunn-Stryker stars and [Fe/H]=-2.12 locus of NGC2419 stars, as expected. 

We have derived $Q$ parameters (Johnson \& Morgan 1953) for a number of main sequence stars in the field, comparing
to the synthetic photometry derived by Majewski (1992) for Gunn-Stryker (1983) spectrophotometric main sequence stars.  Of course, 
this estimation assumes that the field stars are all solar-metallicity dwarf stars, which is almost 
certainly erroneous.  However, as a crude confirmation of the reddening, these assumptions may be adequate.  Restricting 
our analysis to stars on the linear regions of $Q-(B-V)$ space (Figure 4c), we find that
the reddening of the main sequence field stars is $E_{B-V}=0.09 \pm 0.03$, a range which includes the 
estimates of ORS (0.05), Schlegel et al. (0.11) and the reddening implied by the isochrone in Figure 3c.  If the field stars 
are metal-poor, the effect would be to 
push our reddening estimate lower.  Any obscuring dust between the Galactic stars and the cluster would cause an underestimate
of the reddening.  This value for the reddening would represent a minimum and we have therefore utilized 
the higher Schlegel et al. reddening for our analysis.

Our primary motivation for obtaining $U$-band photometry was for the identification of candidate QSO's near the cluster.
The number of background galaxies used as the absolute reference frame for the measurement of proper motion is the limiting factor
in the accuracy of those measures.  In an effort to increase the number of {\it compact, well-centroided} background reference objects 
in our study, we used our UBV photometry to identify any low-probability, well-measured objects that deviated from the stellar
locus in color-color space (c.f. Koo et al. 1986).  We have identified one such object (ID 148, P=21\%) marked on Figure 4 with a star.  This 
will not be used a reference object because it has yet to be spectroscopically confirmed.  This object does, however, lie near the 
reference frame zero point defined in \S5.1.

\section{Cluster Dynamics}

\subsection{Spectroscopic Stars and the Radial Velocity of Pal 13}

The most common radial velocity for Pal 13 quoted in the literature is the Hartwick \& Sargent (1978) value of -27 $\pm$ 30 
km s$^{-1}$.  We have cross-identified the two stars used in that study to our own sample.  Their star 1 is the RR Lyrae 
variable V2, for which we were unable to measure a proper motion.  Based on its variability, location and 
magnitude, it is very likely part of Pal 13.  Their star 2 (listed in Table 
5 as HS2) is almost certainly a member based on its proper motion.
Kulessa \& Lynden-Bell (1992) quoted an unpublished radial velocity measurement of +13 km s$^{-1}$ for Pal 13.  
As no finding charts have ever been published for their radial velocity determination, we can not 
confirm the membership of any stars used in that study.
In Cudworth et al. (1993), we presented a radial velocity for Pal 13 of 
+33 $\pm$ 20 km s$^{-1}$ based upon
six photographic spectra taken by Ruth Peterson of stars with confirmed proper motion membership.  This result has since
been improved upon with 19 high precision velocity measures by Cote (2000) to
a value of $+24.23 \pm 0.45$.  We have confirmed the membership of these stars by comparison to the
proper motion.	Our space velocity calculations below will use this astrometrically confirmed result.

\subsection{Absolute Proper Motion and True Space Velocity}

	The vector point diagram of the Pal 13 sample is shown in Figure 5.  Note the separation of the 
	cluster from other objects in the field (mostly field stars and the few extra-galactic objects).

	To derive absolute proper motions, we assume that galaxies define the absolute zero point of proper 
	motion.  We have a total of 16 centroidable galaxies in our astrometric field.  Two galaxies were removed from this 
	set: one (436) was on the edge of the usable astrometric
	field and a second (952) had very faint images on nearly every plate.  Both of these galaxies also had large proper
	motion errors.
 
	Although the plate constants have removed any measurable systematic magnitude and color dependence from 
	the stellar 
	proper motions, it is possible that the plate constants appropriate for stars might not be
	appropriate for galaxy images that necessarily have different image structure and spectral energy 
	distributions.  Additionally, as the cluster stars used to define the plate constants occupy a narrow
	range in color and the extra-galactic objects mostly reside outside of this range, there may be residual
	uncorrected color effects in the galaxy proper motions.
	Plots 
	of proper motion against magnitude for the galaxies show no systematic trends.  However, plots of 
	galaxy motion against color show a noticeable trend (see Figure 2).

	To correct for this effect, we fit a line to 
	the error-weighted proper motions of the galaxies and defined their net proper motion to be where this 
	line intercepted the average photographic color of stars in the plate constant reference frame 
	($B-V=0.60$).  This effectively sets the zero point of proper motion to that of an extra-galactic object
	with no color difference from the stars used to define the plate constants.  This necessitated the removal
	of two more galaxies from our reference frame (738 and 902) that did not have measured photographic colors
	\footnote{The CCD colors of these two galaxies follow the color-proper motion trend of the other galaxies 
	but have not been included in our solution since the plate constants were derived from photographic colors and 
	magnitudes.  Translation 
	of CCD to
	plate colors is not a simple matter with galaxies since our CCD and photographic photometry methods can produce different
	measures for diffuse non-stellar objects.  If the CCD colors are used for the two faint objects, the proper motion of Pal 13 
	changes to $(\mu_{\alpha cos \delta}, \mu_{\delta}) = (+2.15, +0.46) \pm (0.26, 0.24)$ mas yr$^{-1}$.}.
	Figure 6 shows the vector point diagram of the relative proper motions of the 14 (12) galaxies used in deriving the 
	absolute proper motion of the cluster before (after) the color correction was
	applied.  These motions differ significantly from zero because they are 
	relative to the mean motion of the cluster.
	
	It is important to note that this correction applies only to the {\it galaxies} in our proper 
	motion catalogue.  Figure 2 shows that the {\it stars} have no systematic color trends.  The 
	correction removes the color-related skew of the absolute reference frame to which we are 
	attempting the zero-point tie-in.  The color-corrected motions of the galaxies are listed in 
	Table 6.

	We find the proper motion of the cluster relative to the color-corrected mean of these 12 galaxies is 
	$(\mu_{\alpha cos \delta}, \mu_{\delta}) = (+2.30, +0.27) \pm (0.26, 0.25)$ mas yr$^{-1}$, where the uncertainty is 
	due almost entirely to the galaxy proper motions errors defining the zero point.  
	If the color correction is not applied and the two
	galaxies without measured colors are retained, 
	the proper motion of the cluster is $(\mu_{\alpha cos \delta}, \mu_{\delta})=(+3.41,+0.14)$.  If the QSO candidate identified in 
	\S4.5 (which follows the color-$\mu$ trend) were added to the reference frame, the absolute proper motion
	would be $(\mu_{\alpha cos \delta}, \mu_{\delta}) = (+2.45, -0.56) \pm (0.25, 0.23)$ mas yr$^{-1}$.  This again demonstrates how proper
	motion can be substantially changed by minor revisions to the sample of objects defining the reference frame.

	Using the computational matrices of Johnson \& Soderblom (1987) and the color-corrected proper motion
	without including the candidate QSO, we derive a cluster space velocity of 
	($U,V,W$)=(237,-35,-95) $\pm$ (37,21,23) km s$^{-1}$ relative to the LSR for Pal 13.
	In this configuration, $U$ is 
	the vector toward that Galactic anti-center (reversed from the Johnson \& Soderblom orientation), $V$ is in 
	the direction of Galactic rotation and $W$ is toward the North Galactic Pole.  The solar motion is 
	taken as the Ratnatunga et al. (1989) value of (-11.0,14.0,7.5) and we have used a solar Galactocentric distance of 8.0 kpc.

	Removing the LSR motion (220 km s$^{-1}$) and translating to Pal 13's position produces a
	Galactocentric space velocity of $(\Pi,\Theta,Z)=(258,-154,-95) \pm (24,35,23)$ km s$^{-1}$.  In this 
	left-handed cylindrical system, $\Pi$ is directed outward from the Galactic center, $\Theta$ is in the direction of 
	Galactic rotation, and $Z$ is northward perpendicular to the plane of the Galaxy.  Note that the 
	orientation of our coordinate system has $\Pi$ and $\Theta$ as seen at the cluster, not at the Sun.
	Pal 13 has a large total space velocity of 315 $\pm$ 48 km s$^{-1}$.  This fact, combined with the velocity vector
	described above, suggests that is on a retrograde, eccentric orbit and near
	its perigalactic point.  This point is addressed in greater detail in \S6.

	Without the color correction to the galaxy proper motions, the velocities of Pal 13 are ($U,V,W$)=(342,-86,-156) and 
	$(\Pi,\Theta,Z)=(249,-270,-156)$.  This would give the cluster a total space velocity of 399 km s$^{-1}$, a value
	near the escape velocity of the Milky Way (c.f. Kochanek 1996).

\subsection{Cluster Center and Radial Profile of Pal 13}

It is of interest to measure the structural profile of Pal 13 because the cluster's small size suggests that it 
may be rapidly evaporating stars.  The sparseness of Pal 13 can complicate measurement of structural parameters via a simple starcount
analysis due to the cluster's low contrast against the Galactic foreground.
However, this situation can be ameliorated by selecting likely member stars {\it a priori} via proper motion.

We have fit a circular King (1962) profile to the distribution of stars with probability 
memberships $\ge 50\%$ using a least squares technique.  We measure
a limiting radius ($r_K$)\footnote{This measure is generally referred to in the literature as the ``tidal radius" and abbreviated $r_t$.  
However, the ``tidal radius" of the King profile is not identical to the radius at which the cluster and Galactic potentials are equal.  
Only at perigalacticon are the two measures equal.  To distinguish between the observational measured
radius and the present tidal radius, we refer to the latter at the ``limiting radius", $r_K$.} of 188" $\pm$ 9", a core radius ($r_c$) of 39" $\pm 4"$ and a concentration parameter 
of $c=log(r_K/r_c)=0.7$.
At Pal 13's revised distance, the limiting radius corresponds to a physical cluster 
radius of $24 \pm$ 1 pc.  Figure 7 shows the fit to the 
counts.  We note that fourteen of 119 likely Pal 13 
members in our sample are beyond the limiting radius, including one RR Lyrae variable.  These fourteen stars are 
probably in the process of evaporating or being tidally stripped from Pal 13 (see \S6.4).  Such a signature is seen in numerous
globular clusters (c.f. Grillmair 1998; Leon et al. 2000) and the Carina dSph (Majewski et al. 2000b)

We have also found the center of Pal 13 to be slightly off from the coordinates given in Harris (1996).  By 
centroiding the distribution of probable members, we find the cluster center to be at 
$(\alpha_{J2000.0},\delta_{J2000.0})=(23:06:44.8, 12:46:18)$.

\section{Discussion}

\subsection{The Mass of the Milky Way}

Given the high space velocity of Pal 13, it is natural to ask whether the cluster is bound to the 
Galaxy.  The minimum Galactic mass needed to bind Pal 13 can be derived from its Galactocentric radius and velocity 
by a simple escape velocity calculation.  Using the formulation $M=v_{esc}^2r/2G$, an approximate mass of 
3 x $10^{11} M_{\sun}$ interior to its present position would keep Pal 13 bound.

Estimates of the Milky Way mass have a large degree of uncertainty.  Zaritsky et al. (1989) estimated the mass of the Galaxy 
to be 9-12 x $10^{11} M_{\sun}$ 
if Leo I is bound, 4 x $10^{11} M_{\sun}$ if not.  Cudworth (1990) used local stars to place a minimum Galactic 
mass at 4x$10^{11} M_{\sun}$.  Using the velocity distribution of outer halo objects as revealed by proper motions
and radial velocities, Kulessa \& Lynden-Bell (1992) 
have estimated a Galactic mass as high as $10^{12} M_{\sun}$ while Kochanek (1996) estimated a mass of 
3.9-5.1 x $10^{11} M_{\sun}$ inside 50 kpc,
depending on the status of Leo I.  Most recently, Wilkinson \& Evans (1999) used this method to estimate a Galactic mass of 
5 x $10^{11} M_{\sun}$ inside 50 kpc independent of Leo I.  The proper motions of Pal 13 and other distant halo objects
could precipitate substantial improvements upon these mass estimates (c.f. Wilkinson \& Evans 1999) and this is one
of the ultimate goals of our astrometric survey of outer halo objects.  The 
current estimates of the Milky Way mass are all high enough to bind Pal 13.

\subsection{Orbit and Dynamical History of Pal 13}

The well-established position and distance of Pal 13 in combination with our new space velocity can be used to 
predict the cluster's future orbit or outline its past orbit.  For this purpose, we have used the orbital integration tool and model 
of Johnston et al. (1995, hereafter JSH95).  Figure 8 shows the orbit of Pal 13 integrated 10 Gyr into the 
future.  The rosette pattern that results from its present velocity is fairly typical of outer halo globular 
clusters (c.f. Dinescu et al. 1999, hereafter D99).  Table 7 lists the calculated orbital parameters of Pal 13, based on the true 
space velocity presented in this paper and the JSH95 potential.  Orbital elements were calculated as
in D99 with the small changes that inclination angle ($\Psi$) is calculated at apogalacticon and the period is the
perigalactic period as opposed to the azimuthal period.

As expected from the high velocity, Pal 13
spends most of its orbital time in the distant halo.  It is presently only 70 Myr past 
perigalacticon and spends only 12\% of its 1.1 Gyr orbit this close to the Galactic Center.

The non-color-corrected proper motion of Pal 13 produces a much more radical orbit with apogalacticon at 185 kpc and a 2.6 Gyr
orbital period.  The high velocity
given by the uncorrected proper motion is so extreme that any orbital integration is uniquely sensitive to model 
parameters in a way that other cluster motion integrations are not.  Therefore, the associated orbital parameters are very uncertain because the 
JSH95 model uses an infinite logarithmic potential for the Galactic halo.  Under these circumstances, the apogalacticon and 
eccentricity would represent only lower bounds on Pal 13's orbit.  In either the color-corrected or uncorrected case, Pal 13 is 
on a highly eccentric, retrograde, inclined orbit 
and spends the vast majority of its time in the distant reaches of the Galactic halo.  This point is 
discussed further in \S6.5.

Lynden-Bell \& Lynden-Bell (1995) proposed two dynamical associations of Pal 13 with dwarf spheroidal galaxies and other
globular clusters and made 
predictions for its proper motion supposing such association.  Neither of these proper motions is close to the value we 
have derived for 
Pal 13.  Majewski (1994) suggested a possible association with the Magellanic stream.  Figure 9 shows the orbital
poles of Pal 13 and various proposed ``dynamical families", using the methods detailed in Palma et al. (2000).  We find 
that Pal 13's orbital pole is significantly removed from that of any previously proposed family, including the Magellanic
Stream, the 
Fornax-Leo-Sculptor Stream and the Sagittarius dSph.

Pal 13's association with some of the presently known Galactic 
satellites can also be ruled out by comparing their dynamical 
properties and histories.  Pal 13 has a very high eccentricity.  If Pal 13 had been tidally stripped
from a parent object, that interaction would leave Pal 13 on an orbit that is 10-30\% more or less energetic
than its parent object (Johnston 1998).  As this stripping would happen at or near the pericenter of
the parent object's orbit, this would result in a less eccentric orbit for Pal 13 if its on a less energetic
orbit, more eccentric if on a more energetic orbit.

We have compared Pal 13's dynamical
properties to the five satellite galaxies for which proper motions have been measured.  These are Draco (Scholz \&
Irwin 1994),
Ursa Minor (Schweitzer et al. 2000), Sculptor (Schweitzer et al. 1995), Sagittarius (using the HST proper motion of 
Ibata et al. 2000, which is similar to the ground-based measure of Irwin et al. 1996) and the LMC (an average
of the motions from Jones et al. 1994, Kroupa et al. 1994 and Kroupa \& Bastian 1997)
\footnote{We note that a recent LMC proper motion has been published by Anguita et al. (2000), where the proper motion
has been referenced to three quasars.  We have not included
this result primarily because it is substantially different from the two previous independent measures
and implies a space velocity in excess of the escape velocity of the Milky Way.  Our own experience (Majewski 1992) using quasars
as part of an extra-galactic reference frame suggests problems in the Anguita et al. measures resulting from
differences in the spectral energy distributions of stars and quasars.}.
We have combined these proper motions with radial velocities, distances and positions from 
Mateo (1998) and references therein to derive orbits for the satellite galaxies, and then compared these 
orbits to that of 
Pal 13.  Table 8 lists the proper motions, orbital energy, average total angular momentum, 
perigalacticon, apogalacticon and eccentricity for Pal 13 and the dSph galaxies with measured proper motions.

We find that we can rule out all of the dwarf galaxies with measured proper motions from an association with Pal 13 under the 
scenario of a simple Milky Way-satellite evolution.  
Draco's proper motion implies a velocity too large for it to be bound to the Milky Way.  The 
JSH95 model has an infinite halo that will eventually pull Draco back into the Galaxy.  However, such
a halo is clearly not appropriate if Draco's velocity is as extreme as suggested by its proper
motion.  The large uncertainties do include a number of bound orbits but do not allow a good dynamical comparison
to Pal 13.  The LMC, Sculptor and Ursa Minor have more energetic orbits than Pal 13 but smaller eccentricities.

Sagittarius has 24\% less energy than Pal 13 and a smaller eccentricity.   However, Sagittarius has an angular momentum
approximately 40\% smaller than that of Pal 13, a difference too substantial for a common origin.

These energy and orbital comparisons are, of course, only valid in the case of secular evolution of a two-body Milky Way-dSph 
interaction.  They do not rule out three-body-encounters, e.g. the collision of two satellites in the Galactic halo as
explored by Zhao (1998).

\subsection{Comparing Two Measures of Perigalacticon}

Because Pal 13 has an eccentric orbit, the cluster would seem to be an
interesting object with which to test the assertion that the King profile limiting radius ($r_K$)
reflects the tidal radius of the cluster at perigalacticon  (King 1962; Innanen et al. 1983).  Such an 
argument was used by ORS to speculate that Pal 13 was near perigalacticon.  Oh \& Lin (1992) tested
orbits of $e=0.5$ using a Fokker-Planck code and found no orbital phase dependence for
the limiting radius, which indicates that a cluster's limiting radius should always reflect its perigalactic
tidal radius.  Pal 13 would present a unique observational test of this scenario as 
the cluster will be in the distant halo for a period longer than its relaxation time 
(3 x 10$^8$ years, Gnedin \& Ostriker 1997) and could expand to a larger radius as its tidal radius expands
with declining Galactic potential.

Using the formulation of Innanen et al., our revision of $r_K$, a Pal 13 mass of $5 \times 10^3 M_{\sun}$ 
(based on a more modern M/L ratio to 3.0, Gnedin \& Ostriker 1997) and a Galactic mass of 
$5 \times 10^{11} M_{\sun}$ we estimate the 
perigalacticon of Pal 13 to be 26 $\pm$ 1 kpc, which is significantly larger than
the perigalacticon calculated above from the cluster space motion (11.2 kpc), and is within error bars of the cluster's 
{\it present} distance from the Galaxy.  The inconsistency between the kinematic and tidal estimates of Pal 13's 
perigalacticon is notable and perhaps cautions against assuming that the limiting radius and perigalactic tidal radius
are identical.

\subsection{The Fate of Pal 13}

Pal 13's extremely small size is either the result of having been formed as a small globular
cluster or having been reduced in size by internal or external forces.  Gnedin \& Ostriker (1997)
estimate that the only destructive process that will have major effect upon this particular cluster is evaporation, a mechanism
that will completely destroy the cluster within the next Hubble time\footnote{The possibility of disruption by a dark halo of 
black holes has been addressed by Moore (1993) with a specific analysis of Pal 13.  The conclusion was that Pal 13 does not
show evidence of having been disrupted by such encounters.}.
However, their evaluation assumed a proper
motion for Pal 13 consistent with a kinematical model.  The fact that Pal 13 passes much closer to the Milky Way than previously
suspected and, on closest approach, will not be able to retain the stars in the outer edge of its King profile,
makes it more vulnerable to disk and/or tidal shocking than anticipated in the Gnedin \& Ostriker study.  The possibility
of a more rapid destruction of Pal 13 seems likely.

Clues to substantial mass loss heretofore lie in Pal 13's present stellar population.  Equipartition of energy among cluster
stars means that evaporation and tidal stripping preferentially occur for lower mass stars, while more massive stars, 
particularly binaries, will be left behind in the cluster.  Pal 13's CMD yields two indicators of a disproportionate binary 
content compared to ``normal clusters".  The first is the apparent double subgiant branch, possibly made more evident in 
Pal 13 due to the preferential reduction of non-binary subgiant stars with time.  However, as we have noted, the stars in the ``double
subgiant branch" are not found to be centrally concentrated, as might be expected.

The second clue, of course, is the relatively 
substantial BSS population for a cluster of this mass and miniscule RGB content.  We have already shown that the BSS population is
concentrated to the core showing that mass segregation has likely occurred.  If BSS are understood to be merged binaries then
their high specific frequency in low mass, low concentration clusters like Pal 13 (c.f. Fusi-Pecci 1992) can be understood as the 
result of substantial tidal stripping.  Indeed, Pal 13 is almost a twin of the globular cluster E3 in BSS specific frequency, mass 
and concentration (van den Bergh et al. 1980).  van den Bergh et al. argue these traits to be strong evidence of the death of E3, 
and we believe these arguments apply to Pal 13 as well.  Given the apparent strong past evolution of Pal 13, it is likely that
it is on the fast track to complete destruction by tidal forces.  Rapid evaporation would produce the mass segregation
and preferential lingering of high mass stars, but would not explain the low concentration parameter.  Cluster evaporation elevates
the concentration parameter preceding core collapse (c.f. Elson 1999).  Of course, we have evidence of substantial mass loss in Pal 13
now, as more than 10\% of its members stars are outside the nominal limiting radius (Figure 7) which would be equal
to the tidal radius at Pal 13's present distance.  These stars are too far away from Pal 13's core to be bound
to the cluster at its present Galactocentric distance.  In addition, they appear to follow an $r^{-\alpha}$ distribution
as predited by Johnston et al. (1999, 2000) and confirmed in the case of Carina by Majewski (2000b).  However, the
stars are too few and too far apart for meaningful profile fitting.

\subsection{Implications for Galactic Formation}

The globular cluster system of the Milky Way has long been thought to hold clues to the formation of the Galaxy (c.f. 
Majewski 1999).  It was the lack of an abundance gradient in the halo globular clusters that led to the proposal that part of 
the Galaxy was formed from diverse fragments (Searle \& Zinn 1978).  This has been more recently refined (Zinn 1985; Zinn 1993;
D99) into a scenario in which the Milky Way clusters are divided into three populations - a disk
population of metal-rich clusters, and two halo populations of metal poor clusters.  The ``young halo" 
clusters would represent an accreted portion of the Galaxy while the ``old halo" population would represent the portion formed
during the collapse that eventually formed the disk\footnote{Zinn (1996) divides the former ``old
halo" population into ``metal-poor" and ``blue horizontal branch" populations.  However, as both were likely
formed during the collapse of the Galaxy, the distinction between them is unimportant for the present discussion.  Pal 13's 
classification would be identical in both schemes.}.

The ``young halo" clusters
are characterized by younger age, a second parameter effect in their HB, a mean retrograde rotation, more eccentric orbits and higher 
$Z_{max}$ values.
Pal 13 conforms almost perfectly to this paradigm.  While it has only five horizontal branch 
stars, Pal 13's metallicity implies that they would be very blue in the absence of a second parameter effect.  This is 
clearly not the case.  
Moreover, Pal 13 is a relatively young cluster (according to the MSTO fitting of B97), has a very high $Z_{max}$ and is on a 
retrograde orbit.  An object with this orbit 
would not be consistent with Zinn's old halo population unless it
were extremely metal poor.  In the context of accretion, Pal 13 is unlikely to have 
been accreted from the nearby dwarf galaxies with measured proper motions (\S6.2).  However, it could 
have been accreted from an undiscovered dwarf or one with an unmeasured proper motion.

Clearly, insight into the cluster population of the Milky Way will require the measurement of proper 
motions for more outer halo globular clusters and dwarf galaxies, which we are attempting in our ongoing survey.  
However, Pal 13 does pose an interesting question as it is in a very fortunate position.  It is intrinsically 
faint to the point of being almost unnoticeable on even deep exposures.  It is also near its closest point 
to the Milky Way - the point on its orbit where it will spend the least time.  Our orbital integration shows 
that Pal 13 spends only a small fraction of its time at its present distance or closer to the Galactic Center.
We note that the faint halo globular cluster Pal 12 is also near its perigalactic point (Dinescu et al. 2000).  
It is impossible to determine if such detections are mere coincidence or reflective of a large population of 
faint distant halo globular clusters that are only detectable near perigalacticon.  How many more Pal 13's are in 
the outer halo, too faint and too far away to have been discovered?

The heavens have been thoroughly searched for such objects and the outer halo census includes such faint objects
as Pal 4 ($M_V=-5.82, R_{\sun}=100$ kpc),  Pal 3 ($M_V=-5.59, R_{\sun}=90$ kpc), 
Eridanus ($M_V=-4.90, R_{\sun}=80$ kpc), Pal 14 ($M_V=-4.68, R_{\sun}=72$ kpc)
and AM-1 ($M_V=-4.66, R_{\sun}=129$ kpc) (Harris 1996 and references therein).  Any undiscovered cluster in the 
outer halo would have to be even fainter than these objects or at low Galactic latitude.  We note that nine clusters within 
25 kpc of the Sun are fainter.

Would it be possible for a Pal 13-sized object to be missed at large radii?  Pal 13 is more than a magnitude fainter
than the clusters listed above.  At a distance of 100 kpc, Pal 13
would have only have one or two members brighter than $V=20$ and only 15-16 members brighter than $V=21$.  Such
an object would not have been detected by the POSS.  Thus, the possibility of a number of Pal 13-sized outer 
halo globular clusters remains open.

The Palomar globular clusters were discovered by the POSS, a deliberate and planned effort to search the heavens 
for such objects.  The ongoing SLOAN digital sky survey will extend our reach even further and hopefully give 
insight into whether Pal 13 and Pal 12 are simply rare objects at the tail of the Milky Way globular cluster 
mass distribution or the brightest, nearest representatives of a group of very sparse, distant halo clusters.

\acknowledgements

	The authors wish to thank the National Science Foundation (KMC, MT), the David and Lucile Packard 
	Foundation (SRM, MHS), the Observatories of the Carnegie Institute of Washington in the form of a Carnegie
	Fellowship to SRM at which time this work on Pal 13 began and the Dudley Observatory in the form of the 1991 
	Fullam Award (SRM), for their 
	support of this project.  SRM thanks A. Sandage for helpful discussions and access to the Palomar 200" 
	telescope plate material.  The authors also thank K. Johnston for providing the integration tool for 
	the orbital calculation, C. Palma for providing his orbital pole calculator for Figure 9 and D. Dinescu
	and W. Kunkel for helpful discussions.

\figcaption[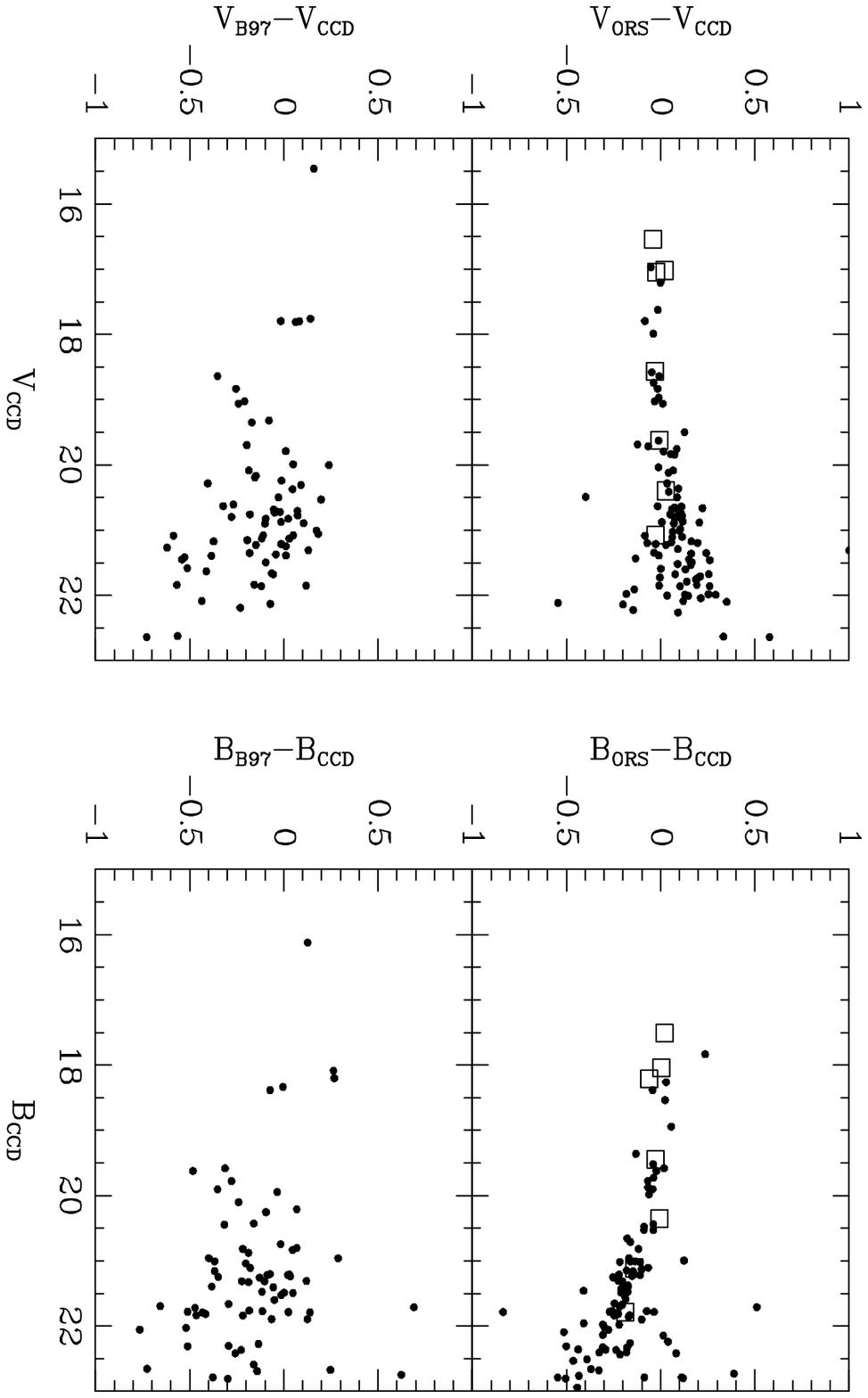]{Comparison of our photometry ($V_{CCD}, B_{CCD}$) to that of ORS and B97.  The open squares in the
ORS comparison mark the photoelectic sequence used in that study.  Note the non-linearity at
the faint end of the ORS comparison.}

\figcaption[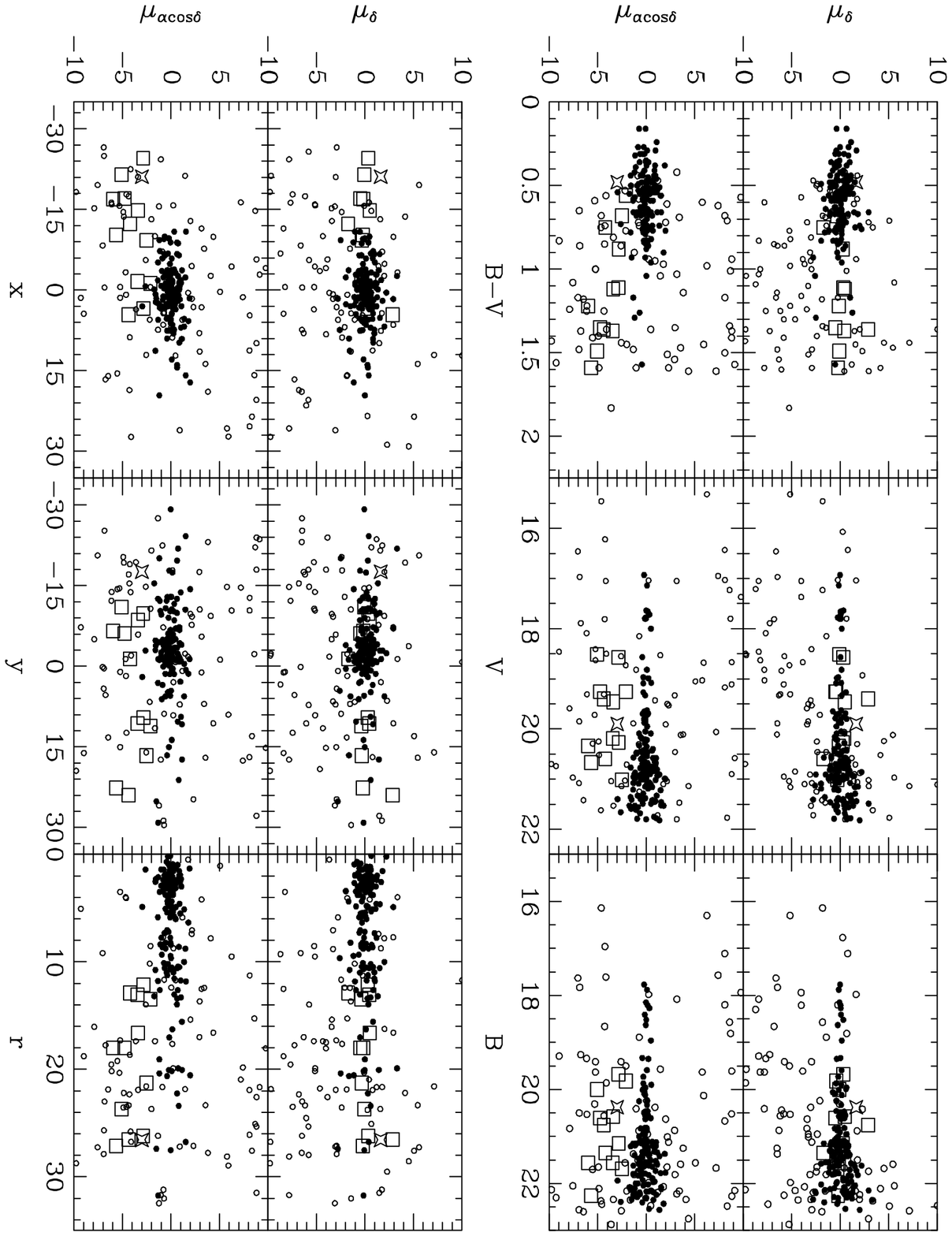]{Plots of proper motion ($\mu_{\alpha cos \delta}, \mu_{\delta}$) against photographic
color, magnitude, position and radius.  Filled circles are stars used to define
the reference frame for the plate constants, open circles are other stars, open boxes are galaxies and the star
is the possible QSO identified in \S4.5.
While the stars show no trend with any parameter, there is a dependence of galaxy proper motion
upon color.  Galaxies removed from the absolute proper motion reference frame in \S5.2 are not
included in these plots.  Note that the photometry illustrated in this figure is from photographic, not the CCD, 
catalogue as plate constants are derived for photographic measures.}

\figcaption[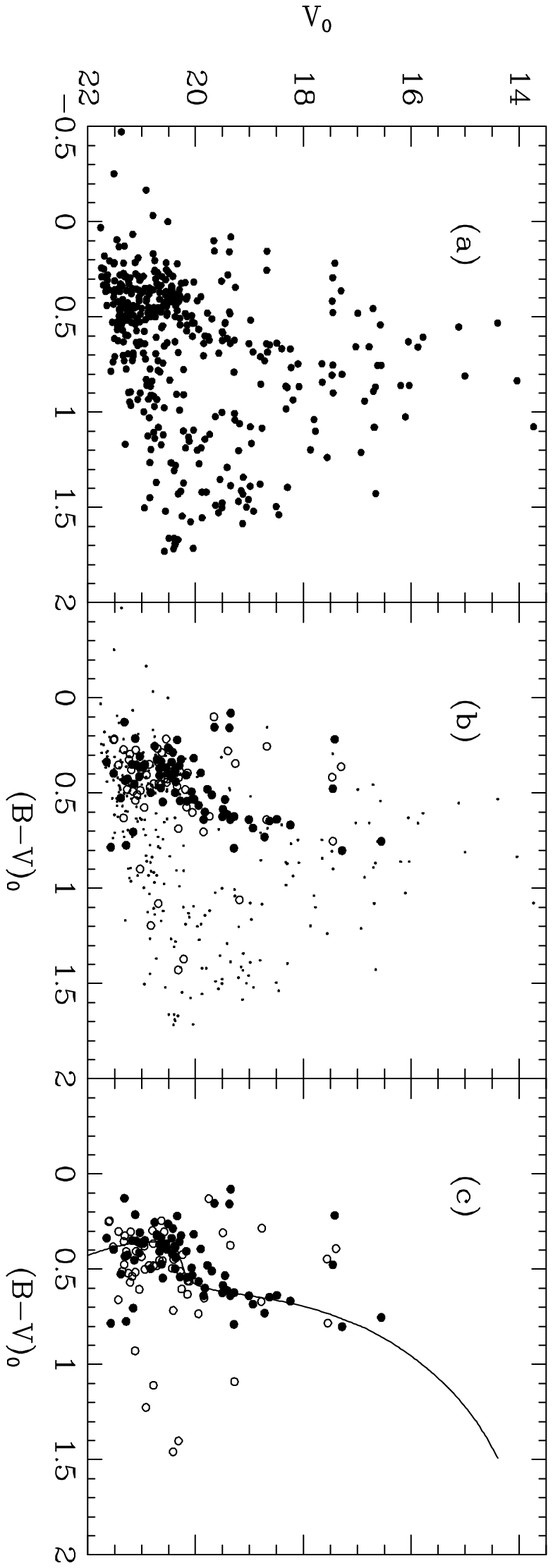]{CCD color-magnitude diagram of Pal 13 as (a) the complete $BV$ data set, (b) with
proper motion data added and (c) cleaned by proper motion with an overlaid isochrone from VandenBerg \& Bergbusch (2000).  Filled
circles are those with membership probabilities $\ge$ 80\%.  Open circles are those with membership
probabilities between 40\% and 80\%.  Dots are objects with either no measured proper motion or a membership
probabilities below 40\%.}

\figcaption[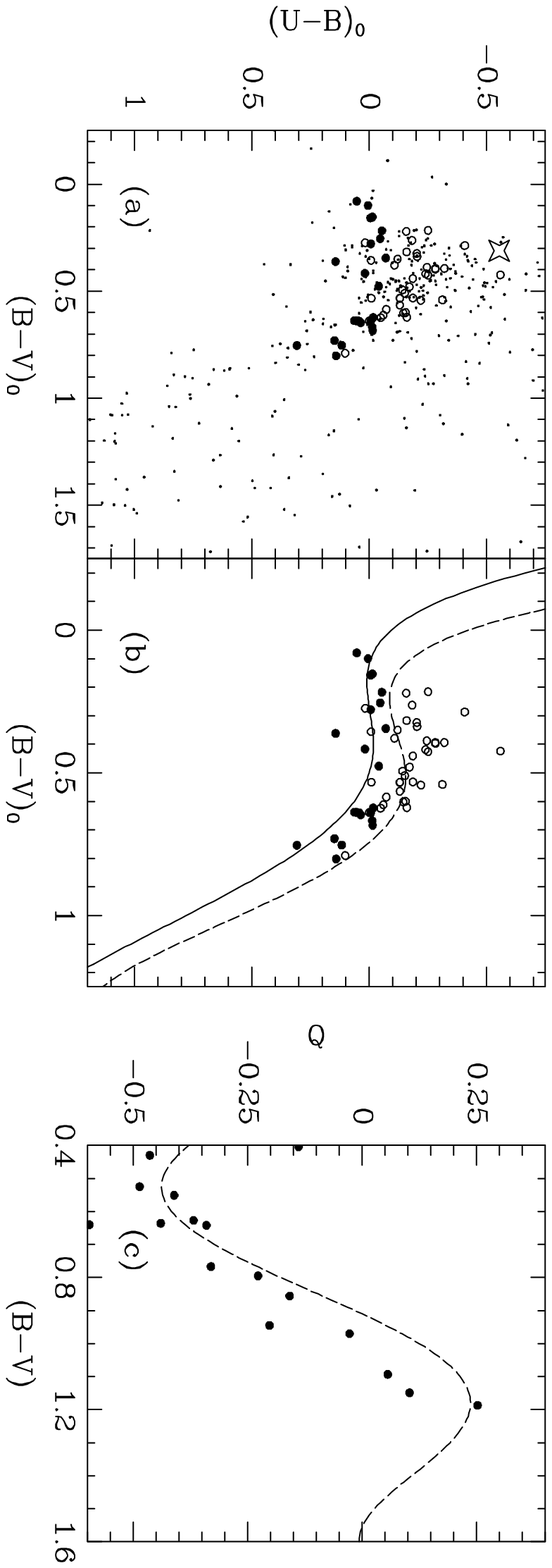]{The two-color diagram of Pal 13 stars.  Part (a) shows all of the data.  Part (b) a selection of Pal 13 
members with good photometry in comparison to a fit to synthetic 
photometry of Gunn-Stryker (1983) standards (solid line) and a fit to the UBV data of NGC 2419 (dashed line).  Open circles are stars with
$P \ge 40\%$, $0.05 < \sigma_U < 0.1$ and filled circles are stars with  $P \ge 40\%$,$\sigma_{U} \le 0.05$.
Part (b) shows stars the high probability members with good $U$ photometry in comparison .  A few stars with no 
measured proper motions that are near the giant branch locus of Figure 4b have been added to increase the range of effective temperature 
in the sample.  Part (c) shows a plot of $Q$ against $B-V$ for the field stars used to estimate foreground reddening.  The reddening vector
is entirely horizontal in this representation.}

\figcaption[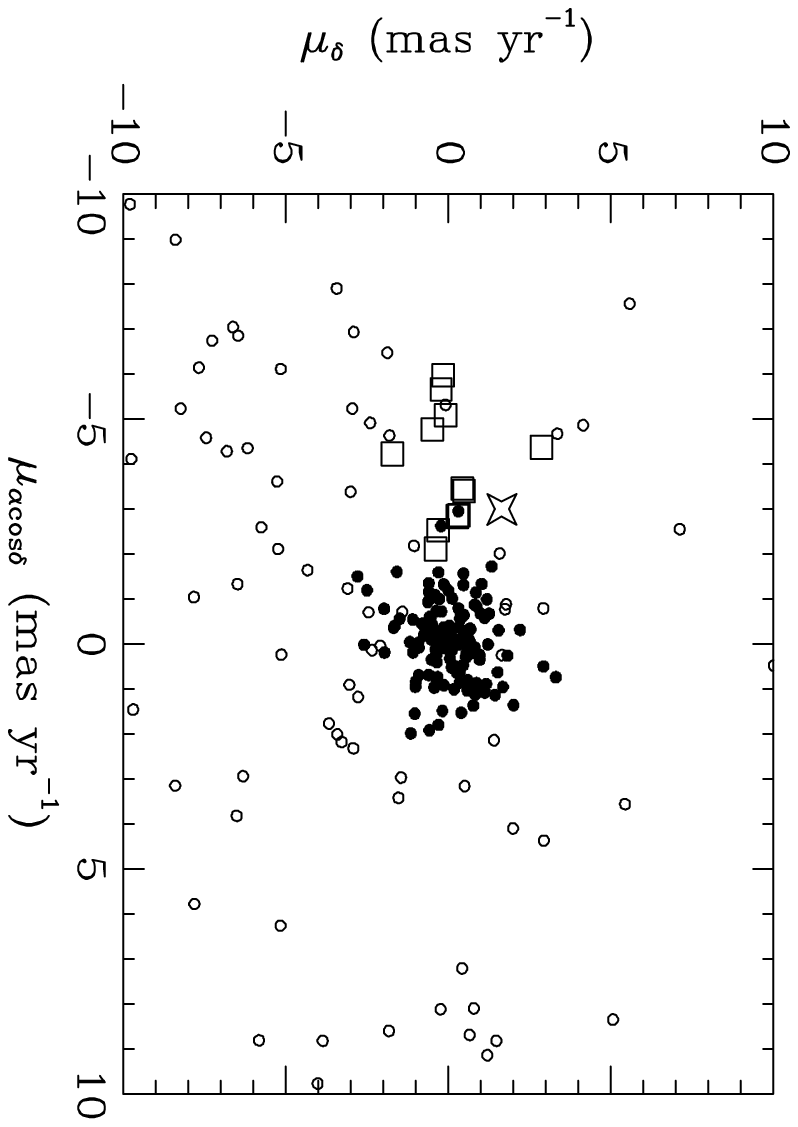]{Vector-point diagram of objects in the Pal 13 astrometric catalogue.  The points represent
the proper motions of individual objects.   The origin of this system is set to the mean proper motion of the globular 
cluster members.  Filled circles are stars used to define
the reference frame for the plate constants, open circles are other stars, open boxes are galaxies, the star is the possible QSO
identified in \S4.5.}

\figcaption[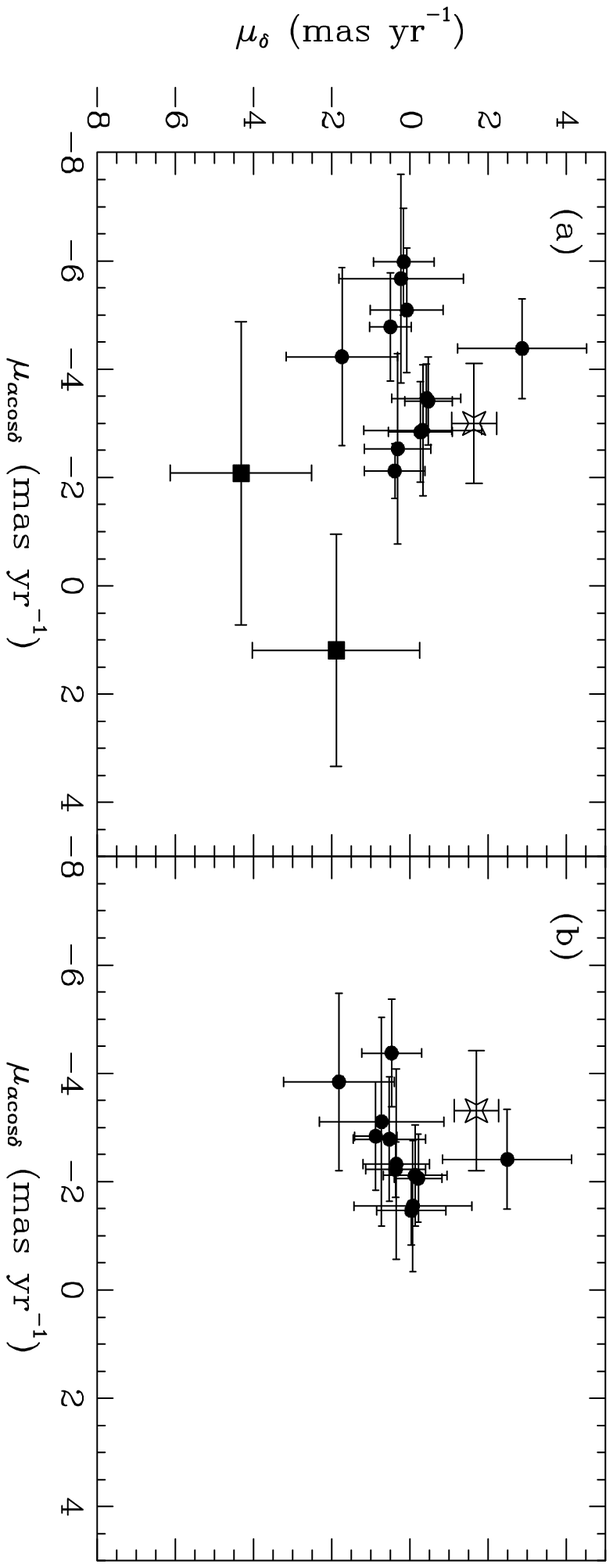]{Vector-point diagram of the extra-galactic objects used to define the absolute frame 
of reference (a) without the color correction and (b) with the color effect described in the text accounted for.  
Note the much tighter clumping of the galaxies in (b).
The two galaxies that did not have measurable colors are symbolized by squares in (a) and do not appear in (b).  The starred
point is the possible QSO identified in \S4.5.}

\figcaption[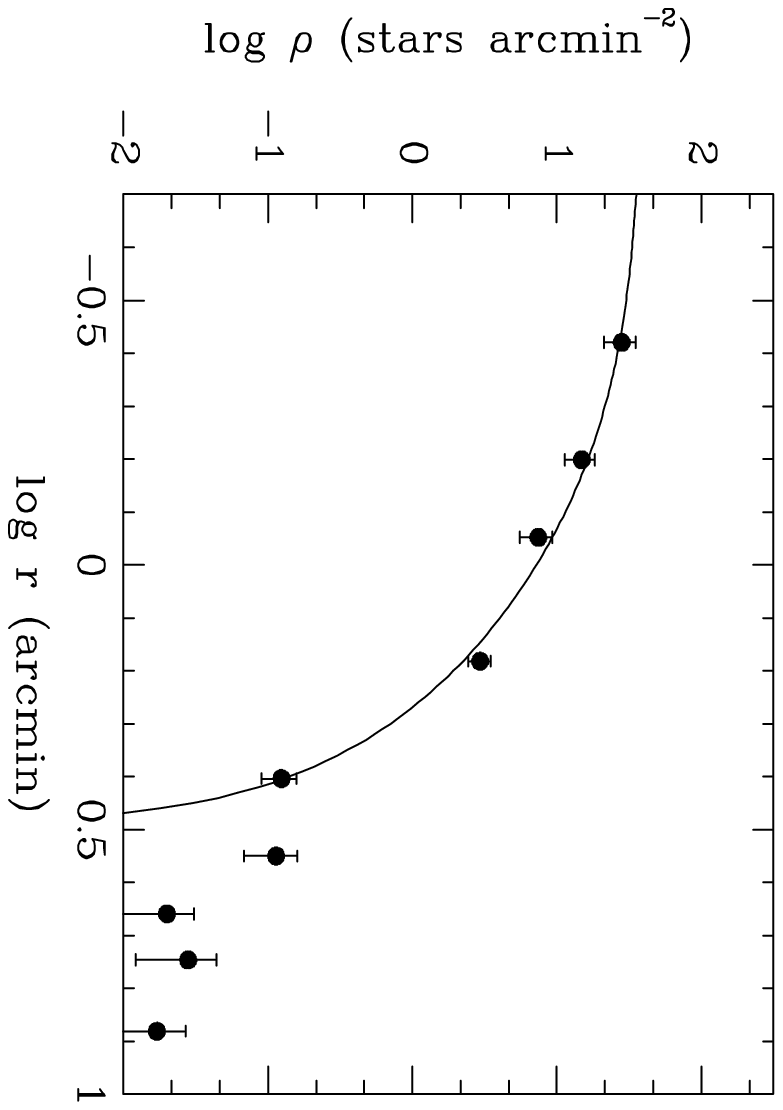]{The radial starcount profile of Pal 13 for stars with membership probabilities
$\ge$ 50\%.  The line is the best fit King profile to the cluster.  Note the member stars outside the classical
limiting radius.}

\figcaption[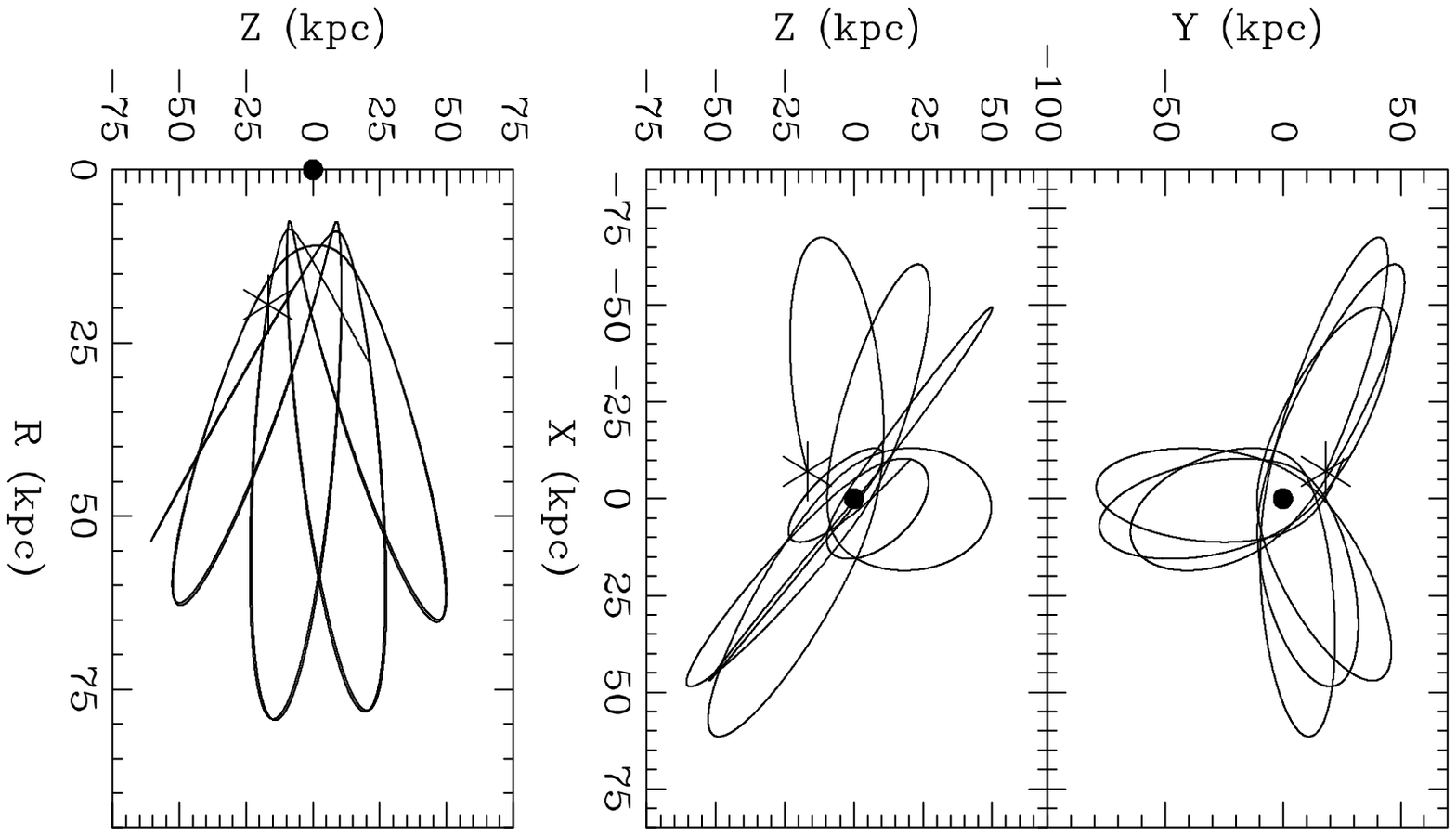]{The orbit of Pal 13 using our new space velocity, position and 
distance for Pal 13 and the model of JSH95.  The top panel represents motion in the plane of the Galaxy while the 
middle panel is perpendicular to the plane.  The bottom panel follows the orbit in a co-rotating X-Z plane.  The star 
marks the current position of Pal 13 while the dot marks the Galactic Center.  Note the familiar 
rosette pattern for outer halo globular clusters as well 
as the proximity of Pal 13 to its perigalacticon.}

\figcaption[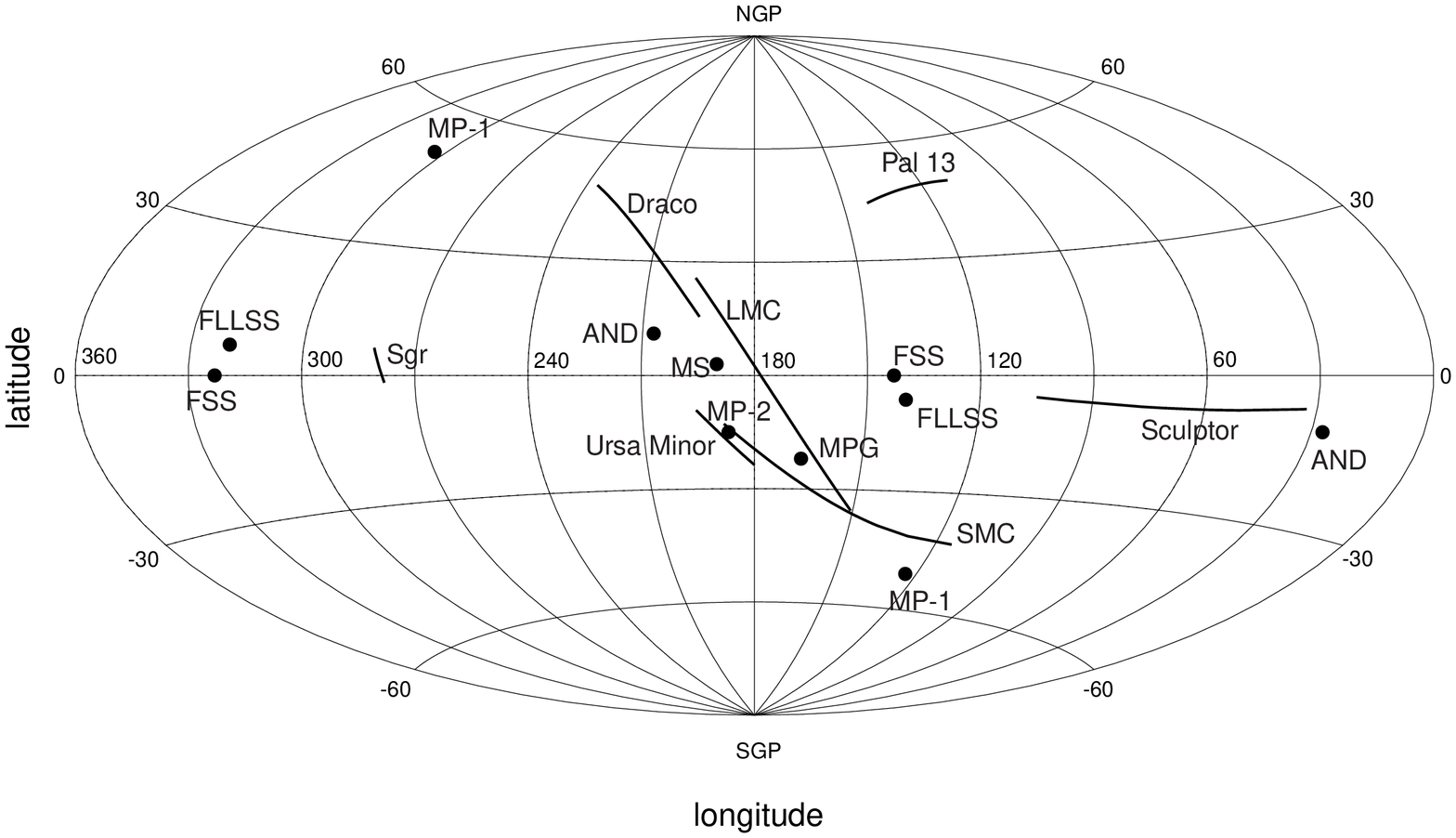]{The orbital pole of Pal 13 in relation to those of the proposed Magellanic Stream 
(MP-1, MP-2, MS, MPG), the Fornax-Leo-Sculptor Stream (FSS, FLLSS), the Andromeda plane (AND) and the dwarf
spheroidal galaxies for which proper motions have been measured (after Palma et al. 2000).}

\clearpage
\begin{deluxetable}{lcccc}
\tablewidth{0 pt}
\tablenum{1}
\tablecaption{Pal 13 Plate Material}
\tablehead{\colhead{Plate No.} & \colhead{Scale ("/mm)} & \colhead{Epoch} & 
\colhead{Emulsion/Filter} & \colhead{Color}}
\startdata
PH-538-B & 11.12 & 1951.76 & 103a-D GG11 & V\\
PH-539-B & 11.12 & 1951.76 & 103a-O GG1  & B\\
PH-812-B & 11.12 & 1953.68 & IIa-O GG1   & B\\
PH-592-S & 11.12 & 1953.77 & 103a-O GG13 & B\\
PH-603-S & 11.12 & 1953.78 & 103a-O GG13 & B\\
PH-865-B & 11.12 & 1953.83 & 103a-O GG13 & B\\
PH-788-S & 11.12 & 1954.65 & 103a-D GG11 & V\\
PH-798-S & 11.12 & 1954.75 & 103a-O GG13 & B\\
PH-813-S & 11.12 & 1954.75 & 103a-D GG11 & V\\
PH-827-S & 11.12 & 1954.75 & 103a-O GG13 & B\\
PH-1062-S & 11.12 & 1955.65 & 103a-D GG11 & V\\
PH-3056-S & 11.12 & 1958.69 & 103a-O GG13 & B\\
PH-3066-S & 11.12 & 1958.69 & 103a-O GG13 & B\\
CD-2925-SM & 10.92 & 1991.52 & IIa-O GG385 & B\\
CD-2931-SM & 10.92 & 1991.53 & IIa-D GG495 & V\\
CD-2932-SM & 10.92 & 1991.53 & IIa-D GG495 & V\\
CD-2939-SM & 10.92 & 1991.53 & IIa-D GG495 & V\\
CD-2947-SM & 10.92 & 1991.54 & IIa-O GG385 & B\\
CD-2954-SM & 10.92 & 1991.54 & IIa-D GG495 & V\\
\enddata
\end{deluxetable}

\clearpage

\begin{deluxetable}{lcc}
\tablewidth{0 pt}
\tablenum{2}
\tablecaption{Probability Parameters for Pal 13}
\tablehead{\colhead{Parameter} & \colhead{Cluster} & \colhead{Field}}
\startdata
    $N$                 & 140  &    86\\
    $\mu_{x0}$ (mas/yr) & 0.0 &  -0.6\\
    $\mu_{y0}$ (mas/yr) & 0.0  &  -2.7\\
 $\sigma_{x0}$ (mas/yr) & 0.6  &   5.8\\
 $\sigma_{y0}$ (mas/yr) & 0.7  &   4.2\\
\enddata
\end{deluxetable}

\clearpage

\begin{deluxetable}{lcccccccc}
\small
\tablewidth{0 pt}
\tablenum{3}
\tablecaption{Photometry Catalogue}
\tablehead{\colhead{ID} & \colhead{ORS} & \colhead{B97} & \colhead{$U$} & \colhead{$\sigma_U$} & 
\colhead{$B$} & \colhead{$\sigma_B$} & \colhead{$V$} & \colhead{$\sigma_V$}}
\startdata
   102 &  \nodata &  \nodata & 20.795 &  0.091 & 20.303 &  0.032 & 19.109 &  0.014 \\
   143 &  \nodata &  \nodata & 21.165 &  0.102 & 20.928 &  0.040 & 19.359 &  0.023 \\
   184 &  \nodata &  \nodata & 21.176 &  0.107 & 20.735 &  0.037 & 19.618 &  0.030 \\
   191 &  \nodata &  \nodata & 21.764 &  0.174 & 21.187 &  0.048 & 19.691 &  0.022 \\
   222 &  \nodata &  \nodata & 20.943 &  0.108 & 20.821 &  0.044 & 20.107 &  0.031 \\
   291 &  \nodata &  \nodata & 22.025 &  0.240 & 21.656 &  0.073 & 20.275 &  0.032 \\
   436 &  \nodata &  \nodata & 21.516 &  0.141 & 22.102 &  0.108 & 20.712 &  0.038 \\
   476 &  \nodata &  \nodata & 21.687 &  0.208 & 21.760 &  0.081 & 20.818 &  0.041 \\
   493 &  \nodata &  \nodata & 21.914 &  0.568 & 22.608 &  0.176 & 20.835 &  0.052 \\
   525 &  \nodata &  \nodata & 23.117 &  0.674 & 22.759 &  0.196 & 20.918 &  0.054 \\
   510 &  \nodata &  \nodata & 21.368 &  0.173 & 21.799 &  0.097 & 20.946 &  0.058 \\
   738 &  \nodata &  \nodata & 22.060 &  0.316 & 22.255 &  0.146 & 21.213 &  0.076 \\
   752 &  \nodata &  \nodata & 21.424 &  0.187 & 22.110 &  0.123 & 21.369 &  0.073 \\
   902 &  \nodata &  \nodata & 21.753 &  0.213 & 22.096 &  0.122 & 21.625 &  0.092 \\
   952 &  \nodata &  \nodata & 21.748 &  0.268 & 22.059 &  0.124 & 21.661 &  0.088 \\
  3671 &  \nodata &  \nodata &  \nodata   &   \nodata  &  \nodata   &   \nodata  &  \nodata   &   \nodata  \\
     1 &  \nodata &  \nodata & 12.681 &  0.016 & 12.695 &  0.014 & 12.144 &  0.011 \\
     2 &  \nodata &  \nodata & 13.670 &  0.015 & 13.652 &  0.011 & 13.016 &  0.008 \\
     4 &  \nodata &  \nodata & 16.368 &  0.019 & 15.261 &  0.013 & 14.074 &  0.012 \\
     5 &  \nodata &  \nodata & 15.804 &  0.020 & 15.326 &  0.010 & 14.381 &  0.007 \\
     6 &  \nodata &  \nodata & 15.503 &  0.015 & 15.381 &  0.009 & 14.739 &  0.007 \\
     8 &  \nodata &  \nodata & 16.695 &  0.015 & 16.267 &  0.009 & 15.347 &  0.007 \\
     7 &  \nodata &   70 & 16.237 &  0.015 & 16.124 &  0.008 & 15.461 &  0.006 \\
    10 &  \nodata &  \nodata & 16.774 &  0.018 & 16.839 &  0.010 & 16.123 &  0.006 \\
    12 &  \nodata &  \nodata & 17.212 &  0.016 & 16.986 &  0.010 & 16.218 &  0.007 \\
\tablebreak
    16 &  \nodata &  \nodata & 18.020 &  0.019 & 17.349 &  0.012 & 16.379 &  0.008 \\
    14 &   31 &  \nodata & 17.167 &  0.021 & 17.135 &  0.011 & 16.395 &  0.007 \\
    18 &  \nodata &  \nodata & 18.718 &  0.021 & 17.585 &  0.010 & 16.450 &  0.007 \\
    19 &   13 &  \nodata & 18.100 &  0.018 & 17.508 &  0.010 & 16.538 &  0.008 \\
    22 &  \nodata &  \nodata & 18.148 &  0.023 & 17.761 &  0.012 & 16.897 &  0.007 \\
    21 &  \nodata &  \nodata & 17.565 &  0.022 & 17.567 &  0.011 & 16.915 &  0.007 \\
    23 &  118 &  \nodata & 18.326 &  0.017 & 17.834 &  0.010 & 16.969 &  0.008 \\
    27 &  \nodata &  \nodata & 19.647 &  0.037 & 18.537 &  0.019 & 17.000 &  0.011 \\
    24 &  \nodata &  \nodata & 18.334 &  0.020 & 17.983 &  0.011 & 17.005 &  0.007 \\
    28 &    6 &  \nodata & 19.386 &  0.026 & 18.211 &  0.012 & 17.021 &  0.009 \\
    30 &  \nodata &  \nodata & 19.355 &  0.026 & 18.223 &  0.012 & 17.033 &  0.009 \\
    26 &  103 &  \nodata & 18.957 &  0.029 & 18.046 &  0.012 & 17.045 &  0.008 \\
    20 &  \nodata &  \nodata & 17.584 &  0.019 & 17.617 &  0.012 & 17.051 &  0.011 \\
    25 &  \nodata &  \nodata & 17.977 &  0.024 & 17.890 &  0.012 & 17.123 &  0.008 \\
    31 &  110 &  \nodata & 19.200 &  0.023 & 18.261 &  0.011 & 17.209 &  0.007 \\
    40 &  \nodata &  \nodata & 19.753 &  0.060 & 18.594 &  0.031 & 17.273 &  0.040 \\
    29 &  \nodata &  \nodata & 17.930 &  0.022 & 17.923 &  0.012 & 17.332 &  0.008 \\
    32 &  \nodata &  \nodata & 18.360 &  0.024 & 18.138 &  0.014 & 17.371 &  0.011 \\
    37 &   72 &  \nodata & 18.757 &  0.023 & 18.537 &  0.011 & 17.625 &  0.007 \\
    36 &  \nodata &  \nodata & 18.337 &  0.020 & 18.115 &  0.014 & 17.643 &  0.009 \\
    33 &  \nodata &   22 & 18.112 &  0.018 & 18.087 &  0.009 & 17.759 &  0.006 \\
    43 &  \nodata &  \nodata & 19.662 &  0.034 & 18.798 &  0.015 & 17.789 &  0.012 \\
    41 &  \nodata &  \nodata & 18.854 &  0.022 & 18.658 &  0.011 & 17.795 &  0.007 \\
    39 &   91 &   23 & 18.421 &  0.018 & 18.383 &  0.010 & 17.796 &  0.007 \\
\enddata
\end{deluxetable}

\clearpage

\begin{deluxetable}{lccccccc}
\small
\tablewidth{0 pt}
\tablenum{4}
\tablecaption{Astrometry Catalogue}
\tablehead{\colhead{ID} & \colhead{$\alpha_{J2000.0}$} & \colhead{$\delta_{J2000.0}$} & 
\colhead{$\mu_x$} & \colhead{$\sigma_x$} & \colhead{$\mu_y$} & \colhead{$\sigma_y$} & 
\colhead{P}}
\startdata
   102 &   23:06:26.06 &   12:44:55.1 &  -2.9 &  1.2  &   0.33 &  1.5  &    -1 \\
   143 &   23:06:28.35 &   12:44:42.6 &  -5.1 &  1.2  &  -0.08 &  0.93 &    -1 \\
   184 &   23:06:33.33 &   12:45:08.4 &  -3.4 &  0.81 &   0.47 &  0.61 &    -1 \\
   191 &   23:06:31.66 &   12:45:36.3 &  -4.8 &  1.0  &  -0.50 &  0.54 &    -1 \\
   222 &   23:06:43.56 &   12:48:43.7 &  -2.1 &  0.51 &  -0.39 &  0.77 &    -1 \\
   291 &   23:06:31.74 &   12:45:30.8 &  -6.0 &  0.99 &  -0.16 &  0.77 &    -1 \\
   436 &   23:06:22.80 &   12:45:46.0 &  -7.9 &  2.9  &  -0.10 &  1.6  &    -1 \\
   476 &   23:06:47.02 &   12:48:26.9 &  -2.8 &  0.93 &   0.27 &  0.82 &    -1 \\
   493 &   23:06:43.27 &   12:48:38.5 &  -3.5 &  0.63 &   0.42 &  0.88 &    -1 \\
   525 &   23:06:36.90 &   12:50:50.0 &  -5.7 &  1.9  &  -0.23 &  1.6  &    -1 \\
   510 &   23:06:37.60 &   12:49:44.1 &  -2.5 &  1.8  &  -0.31 &  0.85 &    -1 \\
   738 &   23:06:37.68 &   12:50:19.9 &  -2.1 &  2.8  &  -4.3  &  1.8  &    -1 \\
   752 &   23:06:35.24 &   12:46:27.8 &  -4.2 &  1.6  &  -1.7  &  1.4  &    -1 \\
   902 &   23:07:01.31 &   12:43:58.4 &   1.2 &  2.1  &  -1.9  &  2.1  &    -1 \\
   952 &   23:06:47.81 &   12:46:01.8 &  -8.5 &  1.0  &   0.13 &  2.4  &    -1 \\
  3671 &   23:06:47.97 &   12:51:04.5 &  -4.4 &  0.92 &   2.9  &  1.7  &    -1 \\
     1 &   23:06:51.39 &   12:45:60.0 &   \nodata &   \nodata &   \nodata &   \nodata &     0 \\
     2 &   23:06:51.07 &   12:45:58.9 &   \nodata &   \nodata &   \nodata &   \nodata &     0 \\
     4 &   23:06:29.01 &   12:44:04.9 &   \nodata &   \nodata &   \nodata &   \nodata &     0 \\
     5 &   23:07:02.77 &   12:47:09.9 &   \nodata &   \nodata &   \nodata &   \nodata &     0 \\
     6 &   23:06:44.73 &   12:44:33.0 &   \nodata &   \nodata &   \nodata &   \nodata &     0 \\
     8 &   23:06:41.09 &   12:44:49.3 &   6.3 &  0.36 &  -5.2  &  0.39 &     0 \\
     7 &   23:06:45.07 &   12:47:02.3 &  -4.6 &  0.28 &  -1.8  &  0.33 &     0 \\
    10 &   23:07:01.88 &   12:45:06.6 &  11.3 &  1.7  &   0.25 &  2.5  &     0 \\
    12 &   23:06:50.89 &   12:43:15.1 &  -4.3 &  0.40 & -24.3  &  0.48 &     0 \\
\tablebreak
    16 &   23:07:01.19 &   12:42:03.1 &   \nodata &   \nodata &   \nodata &   \nodata &     0 \\
    14 &   23:06:52.90 &   12:49:47.2 &   8.1 &  0.20 &   0.78 &  0.53 &     0 \\
    18 &   23:06:38.23 &   12:46:44.1 &  -7.1 &  0.30 &  -6.6  &  0.36 &     0 \\
    19 &   23:06:49.01 &   12:44:46.7 & -20.5 &  0.25 & -19.9  &  0.31 &     0 \\
    22 &   23:06:37.31 &   12:49:26.8 & -0.22 &  0.07 &  -0.02 &  0.10 &    96 \\
    21 &   23:06:45.50 &   12:49:54.3 &   7.4 &  0.31 & -27.4  &  1.0  &     0 \\
    23 &   23:06:43.93 &   12:42:06.9 &  -6.9 &  0.47 &  -6.5  &  0.91 &     0 \\
    27 &   23:06:39.42 &   12:51:11.4 &  42.5 &  0.66 & -12.2  &  1.1  &     0 \\
    24 &   23:06:50.11 &   12:47:14.9 &  0.24 &  0.19 &   1.6  &  0.16 &     0 \\
    28 &   23:06:47.87 &   12:44:23.4 &   8.1 &  0.30 &  -0.25 &  0.62 &     0 \\
    30 &   23:06:50.47 &   12:43:26.1 &  14.4 &  0.35 &  -4.0  &  0.66 &     0 \\
    26 &   23:06:35.86 &   12:46:55.1 &   3.1 &  0.31 &  -8.4  &  0.68 &     0 \\
    20 &   23:06:27.55 &   12:44:12.0 &  -4.2 &  1.8  & -17.1  &  3.0  &     0 \\
    25 &   23:06:54.14 &   12:49:12.2 &  0.08 &  0.06 &  -0.15 &  0.07 &    33 \\
    31 &   23:06:39.36 &   12:47:19.6 & -12.5 &  0.22 &  -8.7  &  0.25 &     0 \\
    40 &   23:06:56.24 &   12:46:17.6 &   8.6 &  0.56 &  -1.8  &  0.76 &     0 \\
    29 &   23:06:41.45 &   12:49:28.5 &   9.8 &  0.25 &  -4.0  &  0.58 &     0 \\
    32 &   23:06:21.30 &   12:47:21.3 &   \nodata &   \nodata &   \nodata &   \nodata &     0 \\
    37 &   23:06:48.51 &   12:46:19.8 &  0.00 &  0.09 &   0.32 &  0.09 &    88 \\
    36 &   23:06:48.84 &   12:41:23.5 & -0.03 &  0.05 &  -0.08 &  0.05 &    66 \\
    33 &   23:06:43.15 &   12:46:47.5 &  0.15 &  0.16 &   0.04 &  0.11 &    97 \\
    43 &   23:06:33.30 &   12:42:40.9 &   8.7 &  1.1  &   0.65 &  0.81 &     0 \\
    41 &   23:06:49.95 &   12:45:28.7 & -0.05 &  0.12 &  -0.25 &  0.12 &    79 \\
    39 &   23:06:43.16 &   12:46:34.3 & -0.15 &  0.11 &   0.08 &  0.10 &    99 \\
\enddata
\end{deluxetable}

\clearpage
\begin{deluxetable}{lcccc}
\tablewidth{0 pt}
\tablenum{5}
\tablecaption{Noteworthy stars in Pal 13}
\tablehead{\colhead{Previous ID} & \colhead{New ID} & \colhead{$V$} &
\colhead{$B-V$} & \colhead{Prob}}
\startdata
ORS-V1    &  38 & 17.81 &  0.53 &   48\%\\
ORS-V2    &  34 & 17.80 &  0.40 &   N/A\\
ORS-V3    &  33 & 17.76 &  0.33 &   97\%\\
ORS-V4    &  36 & 17.64 &  0.35 &   66\%\\
B97-BSS1  & 179 & 19.99 &  0.26 &   81\%\\
B97-BSS2  & 167 & 20.00 &  0.21 &   75\%\\
B97-BSS3  & 306 & 20.76 &  0.40 &   87\%\\
B97-BSS4  & 428 & 20.90 &  0.49 &   23\%\\
B97-BSS5  & 124 & 19.69 &  0.19 &   80\%\\
B97-BSS6  & 126 & 19.71 &  0.27 &   95\%\\
B97-BSS7  & 138 & 19.74 &  0.39 &   79\%\\
BSS8      & 125 & 19.60 &  0.45 &   58\%\\
BSS9      &  79 & 19.02 &  0.36 &   65\%\\
HS2       &  39 & 17.80 &  0.41 &   99\%\\
\enddata
\end{deluxetable}

\begin{deluxetable}{lcccc}
\small
\tablewidth{0 pt}
\tablenum{6}
\tablecaption{Color Corrected Galaxy Motions}
\tablehead{\colhead{ID} & \colhead{$\mu_x$} & \colhead{$\sigma_x$} & \colhead{$\mu_y$} & 
\colhead{$\sigma_y$}}
\startdata
   102 &  -1.55  &  1.21 &   0.08  &  1.51 \\
   143 &  -2.78  &  1.15 &  -0.53  &  0.93 \\
   184 &  -2.06  &  0.81 &   0.21  &  0.61 \\
   191 &  -2.84  &  1.00 &  -0.88  &  0.54 \\
   222 &  -2.22  &  0.51 &  -0.37  &  0.77 \\
   291 &  -4.37  &  0.99 &  -0.47  &  0.77 \\
   476 &  -2.11  &  0.93 &   0.13  &  0.82 \\
   493 &  -1.47  &  0.63 &   0.04  &  0.88 \\
   525 &  -3.11  &  1.93 &  -0.73  &  1.59 \\
   510 &  -2.32  &  1.76 &  -0.35  &  0.85 \\
   752 &  -3.84  &  1.64 &  -1.82  &  1.42 \\
  3671 &  -2.41  &  0.92 &   2.49  &  1.65 \\
\enddata
\end{deluxetable}

\clearpage
\begin{deluxetable}{lc}
\tablewidth{0 pt}
\tablenum{7}
\tablecaption{Orbital Parameters of Pal 13}
\tablehead{\colhead{Parameter} & \colhead{Value}}
\startdata
$R_p$ & 11.2 kpc \\
$R_a$ & 80.8 kpc \\
$Z_{max}$ & 61 kpc \\
$P_{orbit}$& 1.1 Gyr \\
$e$ & 0.76 \\
$\Psi$ &$28^{\circ}$\\
\enddata
\end{deluxetable}

\clearpage
\begin{deluxetable}{lccccccc}
\tablewidth{0 pt}
\tablenum{8}
\tablecaption{Dynamical Comparison of Pal 13 to Dwarf Galaxies}
\tablehead{\colhead{Galaxy} & \colhead{$\mu_{\alpha cos \delta}$} & \colhead{$\mu_{\delta}$} & 
\colhead{E} & \colhead{L} & \colhead{$R_p$} & \colhead{$R_a$} & \colhead{e}\nl
& \colhead{(mas/yr)} & \colhead{(mas/yr)} & \colhead{(km$^2$ s$^{-2}$)} &
\colhead{(kpc km s$^{-1}$)}& \colhead{(kpc)} &\colhead{(kpc)} &}
\startdata
Draco        & $0.6 \pm 0.4$  & $1.1 \pm 0.5$  &$3.034\times10^5$&   N/A           & N/A & N/A  & N/A\\
LMC          &$1.48 \pm 0.23$ &$0.41 \pm 0.37$ &$1.510\times10^5$&   12200         & 34  &98    & 0.49\\
Sagittarius  &$-2.65 \pm 0.08$&$-0.88 \pm 0.08$&$1.137\times10^5$&    5010         & 12  &39    & 0.53 \\
Sculptor     &$0.72 \pm 0.22$ &$-0.06 \pm 0.25$ &$1.652\times10^5$&   18400         & 63  &134   & 0.36 \\
Ursa Minor   &$0.056 \pm 0.078$&$0.078 \pm 0.099$&$1.531\times10^5$&   13700         & 49  &94    & 0.31 \\
Pal 13       &$2.30 \pm 0.26$ &$0.27 \pm 0.25$ &$1.394\times10^5$&    7850         & 11 & 81    & 0.76 \\
\enddata
\end{deluxetable}

\begin{figure}
\plotone{Siegel.P13fig1.eps}
\end{figure}
\begin{figure}
\plotone{Siegel.P13fig2.eps}
\end{figure}
\begin{figure}
\plotone{Siegel.P13fig3.eps}
\end{figure}
\begin{figure}
\plotone{Siegel.P13fig4.eps}
\end{figure}
\begin{figure}
\plotone{Siegel.P13fig5.eps}
\end{figure}
\begin{figure}
\plotone{Siegel.P13fig6.eps}
\end{figure}
\begin{figure}
\plotone{Siegel.P13fig7.eps}
\end{figure}
\begin{figure}
\plotone{Siegel.P13fig8.eps}
\end{figure}
\begin{figure}
\plotfiddle{Siegel.P13fig9.eps}{11in}{0.}{80.}{80.}{-245}{185}
\end{figure}

\end{document}